\documentclass[11pt]{article}

\def\tr{\;{\rm tr}\;}

\def\bra{\langle}   \def\ket{\rangle}

\def\implies{\Rightarrow}
\def\vs#1{\vspace{#1mm}}
\def\fl{\flushleft}

\newcommand{\tl}[1]{\tilde{#1}}

\newcommand{\dd}[2]{\frac {\partial #1}{\partial #2}}
\newcommand{\pdr}{\partial}
\newcommand{\grad}{\nabla}

\newcommand{\beq}{\begin{eqnarray}}
\newcommand{\eeq}{\end{eqnarray}}

\newcommand{\half}{\frac{1}{2}}

\newcommand{\ov}[1]{\frac{1}{#1}}
\newcommand{\fr}[2]{\frac{#1}{#2}}

\newcommand{\N}{\frac{1}{N}}
\newcommand{\Nsq}{\frac{1}{N^2}}

      \def\gb{\beta}   \def\g{\gamma}       \def\G{\Gamma}
\def\gd{\delta}      \def\D{\Delta}  \def\eps{\epsilon} 
          \def\la{\lambda}      \def\La{\Lambda}
                      \def\vphi{\varphi}
             
     \def\si{\sigma}  \def\Si{\Sigma}     \newcommand{\vsi}{\varsigma}

      \def\Om{\Omega}  

\newcommand{\B}{B} 

\usepackage{hyperref}
\usepackage{graphics}
\usepackage{graphicx}
\usepackage{epsf}
\input{epsf}

\begin{document}

\begin{titlepage}

\title{\normalsize \hfill ITP-UU-07/3  \\ \hfill SPIN-07/3
\\ \hfill {\tt hep-th/0701102v2}\\ \vskip 0mm \Large\bf
Naturalness via scale invariance and non-trivial UV fixed points in
a 4d $O(N)$ scalar field model in the large-$N$ limit}

\author{Govind S. Krishnaswami}
\date{\normalsize Institute for Theoretical Physics \& Spinoza Institute \\
Utrecht University, Postbus 80.195, 3508 TD, Utrecht, The
Netherlands
\smallskip \\ e-mail: \tt g.s.krishnaswami@phys.uu.nl \\ March 31, 2007 } 

\maketitle

\begin{quotation} \noindent {\large\bf Abstract } \medskip \\

We try to use scale-invariance and the $1/N$ expansion to
construct a non-trivial $4$d $O(N)$ scalar field model with
controlled UV behavior and naturally light scalar excitations. The
principle is to fix interactions at each order in $1/N$
by requiring the effective action for arbitrary background fields to be scale-invariant.
We find a line of non-trivial UV fixed-points in the large-$N$ limit, parameterized by a
dimensionless coupling. Nether action nor measure is scale invariant, but the effective action is.
Scale invariance makes it natural to
set a mass deformation to zero. The model has phases where $O(N)$
invariance is unbroken or spontaneously broken. Masses of the lightest excitations above the
unbroken vacuum are found. Slowly varying quantum fluctuations
are incorporated at order $1/N$. We find the $1/N$ correction
to the potential, beta function of mass and anomalous
dimensions of fields that preserve a line of fixed points for constant backgrounds.


\end{quotation}

PACS: 11.10.Gh, 
11.15.Pg, 
14.80.Cp, 
11.25.Hf. 

Keywords: non-trivial fixed point, $1/N$ expansion, scale
invariance, naturalness, Higgs particle, fine tuning.

\thispagestyle{empty}

\end{titlepage}




\small

\tableofcontents

\clearpage

\normalsize

\section{Introduction}
\label{s-intro}

We investigate the naturalness concept of 't Hooft
\cite{tHooft-naturalness} applied to $4$d $O(N)$
scalar fields. If there are scalar particles very light
compared to the Planck mass, it must be due to a symmetry. We observe that one
non-trivial scale-invariant RG fixed point in {\em quantum} scalar field theory
would be enough to make small masses natural. For, setting masses to
zero would buy us symmetry under scale transformations. We try to implement this idea by
developing a method due to Rajeev \cite{rajeev-non-triv-fp} for
constructing a fixed point in the large-$N$ limit.

\subsection{Background and motivations}
\label{s-bkgrnd-motivations}

Many discussions of $4$d quantum field theory begin with massive $\la \phi^4$ theory. However, this model
most likely does not have a non-trivial continuum limit
\cite{luscher-weisz,kuti-lin-shen,bender-triviality,rigorous-triviality}. Our first motivation
is to answer the question: `Can one construct a non-trivial $4$d scalar field theory?'

Our second motivation comes from particle physics.
The 2004 Nobel prize in physics for the discovery of asymptotic
freedom in QCD has reminded us about the physical importance of
quantum field theories with well-controlled ultraviolet
behavior. Indeed, Yang-Mills theories, which have a gaussian
high energy fixed point are at the heart of our best models for
both the strong and weak interactions. In equilibrium
statistical mechanics of $O(N)$ magnets, the gaussian fixed point controls high
energy behavior while the lower energy dynamics is governed by
a crossover to the non-trivial Wilson-Fisher fixed point
\cite{wilson-kogut}.

However, the situation in $4$d massive $\la \phi^4$
theory, which is the simplest and currently favored (but
experimentally unconfirmed) model for W and Z mass generation, is
less satisfactory for two reasons. First, $\la \phi^4$ theory
is based on the gaussian infrared fixed point,
but does not flow to any UV fixed point to control the high
energy behavior. Perturbatively, interactions become very
strong at a finite energy $\sim m \exp{[\fr{16\pi^2}{3\la}]}$
where $m,\la$ are the parameters of
the model in the infrared (Landau pole). This is in contrast
with asymptotically free theories or theories based on an
interacting UV fixed point which might (at least in principle)
be defined in the UV. Both numerical
\cite{luscher-weisz,kuti-lin-shen} and
analytical\cite{rigorous-triviality,bender-triviality}
calculations suggest that in the absence of a UV cutoff, the
theory is `trivial'\footnote{The renormalized coupling constant
vanishes identically and correlations satisfy Wick's formula.}.
Unfortunately, the non-trivial Wilson-Fisher fixed point in
quartic scalar field theory in $4-\eps$ dimensions merges with
the gaussian fixed point in 4d. As a practical matter, the lack
of a UV fixed point in $\la \phi^4$ theory does not prevent us
from using it as an effective theory with a cutoff or as a
perturbatively defined model like QED, over a range of
relatively low energies.

The second problem with $4d$ $\la \phi^4$ theory, is the
naturalness problem\footnote{See appendix
\ref{a-eg-naturalness} for an explanation
with examples.}. In QED, a small electron mass
(compared to $m_{\rm Planck}$) is natural because setting it
equal to zero gives QED an additional symmetry, chiral
symmetry, which is not broken by quantum effects. On the other
hand, in massive $\la \phi^4$ theory, setting $m=0$ makes the
classical theory free of any dimensional parameter. But
scale-invariance is broken in the quantum theory due to the
absence of any regularization and renormalization scheme that
preserves scale-invariance, there is a scale anomaly. In the
absence of any symmetry to explain a small scalar mass in the
quantum theory, naturalness suggests that the lightest scalar
excitation, the Higgs particle should have a mass of order the
Planck mass. A very large Higgs mass, however, leads to other
problems since (perturbative, essentially tree level) unitarity
would then be violated \cite{pert-unit-bound}. The perturbative
unitarity bound from a partial wave analysis of scalar exchange
in W-boson scattering is estimated to be of the order of $1$
TeV. Moreover, the likely triviality of the continuum theory
implies a `triviality bound' on the Higgs mass, which is also
of the same order \cite{triviality-bound}. There does not seem
to be any non-perturbative cure for these problems, they also
arise on the lattice \cite{luscher-weisz,kuti-lin-shen} and in
other analytical approaches
\cite{bender-triviality,rigorous-triviality}. 

Another issue is that the $1$-loop correction to the square of
the bare Higgs mass is quadratic in the momentum cutoff and
leads to the fine-tuning problem. If a large cutoff
is to be maintained, either the effective Higgs mass is of the
order of the cutoff or the bare mass must be fine-tuned to
cancel the radiative correction (at each order). We already
mentioned the difficulties with a very large Higgs mass.
However, the fine-tuning problem needs to be treated with some
care. In a renormalizable QFT (eg. $\la
\phi^4$), the use of a cutoff as a regulator is a matter of
convenience. The quadratic divergence is absent in dimensional
or $\zeta$-function regularization or Epstein-Glazer
renormalization. In the end of a calculation, one sends all
regulators to their limiting values before making physical
conclusions. From this standpoint, a quadratically divergent
self energy is not a deficiency of the model.

On the other hand, one may be of the opinion that the QFTs of
particle physics are effective field theories that come with a physical large
momentum cutoff at either the Planck or other scale where new
effects render the standard model inaccurate. This is
similar to ones attitude in some (but not all) condensed matter
physics contexts, where the crystal lattice is physical and not
merely a convenient regulator. In this view,
renormalizability is expendable. The cutoff is not to be sent
to infinity, and the fine tuning problem mentioned above is
indeed present.

In QCD or other renormalizable models that are based on a UV
fixed point (gaussian or not) to control high energy behavior,
there is no need for the `effective field theory-physical cutoff'
view point. For, these models self-consistently predict low
energy behavior irrespective of what the physics beyond the
standard model may be or at what scale it may kick in. But in
models such as $\la \phi^4$, which (despite being
renormalizable) seem to be non-trivially defined
only in the presence of a cutoff, the latter viewpoint cannot
be ignored. This leads us to an important distinction between
the naturalness concept and fine tuning. The former requires
any model to have an extra symmetry to explain a small parameter.
But fine tuning, for its very definition, needs the model to have a
physical cutoff. Fine tuning is potentially an issue for non-renormalizable
models or models which need a cutoff in order to be
non-trivial (eg. $\la \phi^4$).

There is an empirical relation between naturalness and degree
of fine tuning. Suppose we dogmatically insist on
working with a cutoff even though our model may be
renormalizable. Assume also that the model is natural, i.e.
gains a symmetry if masses are set to zero. Then it is often
the case that radiative corrections are `protected by the
symmetry': self energies are only logarithmically divergent
rather than as a power of the cutoff. An example is QED. Due
to chiral symmetry, the renormalization of the electron
mass$^2$ is not quadratically divergent as one might naively
expect, but only logarithmically divergent (the quadratic
divergence persists in scalar QED, which does not have chiral symmetry).

Our conclusions are (1) it is worth looking for a non-trivial
scalar field model based on a UV fixed point;
(2) it is worthwhile to look for a symmetry
that would ensure naturally light scalar excitations and (3) If
(1) and (2) are achieved, it is less important to worry about
fine-tuning in the presence of a cutoff.

Despite our conceptual criticisms of $\la \phi^4$ theory, as
long as a light Higgs is discovered, it will
be possible to use the model to predict scattering amplitudes at
relatively low energies. Thus it remains the default mechanism for
giving masses to the weak gauge bosons. It may turn out to be an
effective description of some more intricate framework devised by
nature.

Many alternatives have been proposed. Most studied is the
use of supersymmetry to ensure light scalars \cite{SUSY-higgs}. The
challenge here is to break SUSY without introducing new naturalness
problems. Technicolor models try to realize scalars as fermion composites
\cite{technicolor}. Little Higgs models \cite{little-higgs} attempt to use the
Nambu-Goldstone mechanism in a novel way. In the
Coleman-Weinberg mechanism \cite{coleman-weinberg}, classical
conformal invariance of massless $\la \phi^4$ theory is used to
explain small scalar masses. But this is not a naturalness
explanation in the strict sense since conformal invariance is broken
quantum mechanically. An idea to use scale invariance in a manner
analogous to SUSY and soft SUSY breaking,
has been proposed \cite{Bardeen:1995kv}. Other approaches include
higher derivative models to expel one-particle scalar excitations as
asymptotic states \cite{slavnov}; and the possibility of the
gaussian fixed point being UV with respect to some non-polynomial
potentials \cite{halpern-huang}.

Richter \cite{richter} has criticized naturalness. However,
our definition of naturalness is not
quite the same as the one he uses (in particular, a quadratic
divergence in self-energy is not unnatural by our
definition). Moreover, his criticisms seem to have more to do
with the large number of parameters in the MSSM, than with
naturalness as we understand it. We hope the examples in
appendix \ref{a-eg-naturalness} will help to improve the dismal
score he gives the concept! Finally, one hopes that experiments at
the LHC will add empirical discrimination to discussions
on naturalness and electro-weak symmetry-breaking.


\subsection{Reasoning and summary}
\label{s-summary}

Motivated by these considerations, we investigate whether there is
any resolution to the difficulties of $\la \phi^4$ theory that does
not require adding new parameters or degrees of freedom beyond those
of the standard model\footnote{In this paper we work in an approximation where gauge and Yukawa couplings vanish.}.
Can we construct an
interacting scalar field theory in $4$d, which has good UV
behavior and supports naturally light scalar excitations? We argue
that one possible scenario is that such a model should be built
around a nontrivial UV fixed point. Existing work does not indicate
(indeed, almost rules out) a non-trivial fixed point in the
neighborhood of the gaussian fixed point\footnote{Halpern and Huang
\cite{halpern-huang} have argued that there may be potentials with
respect to which the gaussian fixed point is UV. But this scenario
is quite different from what we propose.}, so we will look
farther afield.

The two key concepts of this paper are (1) naturally light scalars
from scaling symmetry in the quantum theory and (2) non-trivial
quantum effective actions via cancelation between `action' and
`quantum fluctuations'.

To find non-trivial fixed points far from the gaussian one, we
give up thinking of a QFT as being defined by a classical
action. The reason the classical action is a useful concept
is that it provides a first approximation to the quantum effective
action in the $\hbar \to 0$ limit. Indeed, the contribution of the
path integral measure is suppressed in this limit. By contrast, in
the zeroth order of the approximation method we propose, both the
`action' and the quantum fluctuations from the `measure' are
comparable. In fact, both are strictly infinite prior to
regularization and neither is scale invariant if regulated. However,
their combined effect at the zeroth order of our scheme is to produce a finite $1$-parameter family of
scale invariant (quantum) effective actions when regulators are
removed. It is the latter that is physically observable and defines
the theory. Such a scenario is not amenable to analysis via the loop
expansion, traditional perturbation theory or a weak field expansion
in powers of the scalar field. In particular, we give
up perturbative renormalizability, but require it in
a non-perturbative sense so as to ensure predictive power. We hope
this is excusable since despite its great success elsewhere,
insisting on perturbative renormalizability in scalar field theory
leads to a model that is unnatural, lacking in good UV behavior and
most likely trivial.


To embark on the formidable task of constructing a non-trivial fixed
point far from the gaussian one, it will help to have a small
parameter to expand in. We investigate whether there is a $4$d
scale-invariant interacting $O(N)$ scalar field theory in the $1/N$
expansion. The case of interest in particle physics is $N = 4$,
since in the absence of gauge and Yukawa couplings, the symmetry
group of the standard model is $O(4)$. Our approach is inspired by work on Yang-Mills theory,
where we have learned that the theory has two different
``classical'' limits in which quantum fluctuations of some
observables are small. Both the $\hbar \to 0$ and $N \to \infty$
limits (holding the other parameter fixed) can be used as starting
points for studying the full quantum theory. Might the large-$N$
limit be of use in constructing a scale-invariant $4$d  scalar field
theory? This possibility was pointed out by Rajeev
\cite{rajeev-non-triv-fp}.

If scale-invariance could be maintained even after including quantum
effects, then such a scale-invariant model would define a fixed
point. The way we achieve this, is to pick an action which
is not scale-invariant, so as to cancel the `scale anomaly' from
quantum fluctuations. In the large-$N$ limit, we actually find a
line of fixed points\footnote{A line of fixed points has appeared
previously in 4d QFT. In ${\cal N}=4$ super Yang-Mills theory,
$g^2_{YM}$ is believed to parameterize a line of {\em quantum} fixed
points. In a sense, things are simpler in ${\cal N}=4$ SYM since
$g^2_{YM}=0$ is the trivial fixed point and the line of fixed points
may be studied perturbatively.} parameterized by $\la$. $\la$ is the
dimensionless coupling constant of a $\la \phi^4$-type term which is
marginally irrelevant when considered around the gaussian
fixed-point but whose beta function vanishes in the large-$N$ limit
when considered around the non-trivial fixed point. In the spirit of
Wilsonian renormalization, we should study the most relevant
deformation of the line of fixed points. This corresponds to a mass
term. Setting mass to zero should be natural, since we would gain
scale-invariance by doing so. This line of scale-invariant theories
would be UV fixed points with respect to the mass deformation and
thus ensure good high energy behavior. For naturally light scalars
via scaling symmetry, it is sufficient to have one fixed point. It
may well be that when $1/N$ corrections are included, scale
invariance can be maintained only for one\footnote{There may be no
non-trivial fixed point when $1/N$ corrections are incorporated,
then our scenario would fail.} value of $\la = \la_0$. This would be
acceptable, since $m=0, \la = \la_0$ would be natural due to scale
invariance at that point. It would be a bonus if scale-invariance
reduces the degree of divergence of self energy in the presence of a
cutoff, just as chiral symmetry does in QED. But as a matter of
principle, this is not necessary as long as the model is
renormalizable and non-trivial when regulators are removed. Even
more importantly, we would like locality, causality and unitarity.
Locality is likely to be a subtle issue. We must also investigate the low energy
behavior as we flow away from one of the UV fixed points along a
mass deformation.

Our model is to be constructed order by order in the $1/N$ expansion
by requiring that the effective action for arbitrary backgrounds be
scale-invariant. The physical output of this procedure is the
effective action, not the classical action. Our starting point is a
$4$d $N+1$ component\footnote{It is convenient to work with $N+1$
components $\phi_0, \phi_1, \cdots \phi_N$ since we will integrate
out all but $\phi_0$.} Euclidean scalar field model whose partition function
can be written formally in terms of an $O(N+1)$ invariant potential
$V$
    \beq
    Z = \int [D\phi] \exp{\bigg[ -\ov{\hbar} \int d^4x \bigg\{ \half |\grad \phi_i|^2 + N
        V(\fr{|\phi|^2}{N})  \bigg\} \bigg]}.
    \eeq
It is analyzed by introducing an additional field $\si$ via
a Laplace transform so that the action is quadratic in $\phi_i$.
$\si(x)$ is the Laplace conjugate\footnote{$\si$ is valued on a
contour $\cal C$ from $-i \infty$ to $i \infty$. Within our
approximations, $\cal C$ must avoid ${\bf R}^-$.} of the
$O(N+1)$ singlet $\eta = \fr{\phi_i \phi_i}{N}$ (section
\ref{s-lagrangian-change-of-var}). The reason to work with $\si$
instead of $\phi_i$ is that in the large-$N$ limit holding $\hbar
\ne 0$ fixed, $\si$ has small quantum fluctuations, while $\phi_i$
continue to have large fluctuations. The functional integral over
all but one of the $\phi_i$ is performed, leaving only $b =
\fr{\phi_0}{\sqrt{N}}$ and $\si$ as dynamical fields. Keeping $b$
facilitates discussion of $O(N+1)$ symmetry breaking.

The goal of a QFT is the determination of the (quantum) effective action.
In section \ref{s-cl-eff-ac-arbit-bkgrnd} we show that in the
large-$N$ limit holding $\hbar$ fixed, there is a one parameter
family of finite and scale-invariant effective actions when
regulators are removed. For background fields $B(x),\Si(x)$, it
is obtained in an expansion around a constant background
$\Si_o$. In zeta-function regularization,
    \beq
        \Gamma_0(\B,\Si) = \half \int d^4x \bigg[
            (\grad \B)^2 + \Si \B^2 + \la \Si^2 - \fr{\hbar}{16\pi^2}
            \bigg\{(\Si - \Si_o) \Pi(\Delta) (\Si - \Si_o)
         + {\cal O}(\Si - \Si_o)^3 \bigg\} \bigg].
    \eeq
Here $\D = - \grad^2/\Si_o$ and $\Pi(\D)$ is a specific function determined in appendix
\ref{a-derivative-exp-of-tr-log}. $\Si_o$ is arbitrary\footnote{When $\Si_o$ is a non-positive real
number there are further divergences, which we have not yet treated.},
it is {\em not} a parameter of the theory. It appears merely because we are
studying the theory around a constant background field.
$\G_0(\B,\Si)$ depends on the regularization scheme. For instance,
to relate two schemes for constant backgrounds, $\la$ must be
shifted by a finite additive constant which we have determined for
zeta function, cutoff (sec. \ref{s-mom-cut-off-regularization}) and
dimensional regularization (appendix \ref{a-dim-reg}). The principle
that the effective action be scale-free leaves the dimensionless
coupling $\la$ undetermined, and it parameterizes a line of fixed
points. In section \ref{s-cancel-scale-anomaly} we show that $\G_0$
is free of scale anomalies. Effective actions in QFT usually involve
proper vertices of all orders and are very complicated, ours is no
exception. What is perhaps unusual is that unlike in the zeroth
order of the loop expansion of $\la \phi^4$ theory, the effective
action of our model in the zeroth order of the $1/N$ expansion
already involves vertices of all orders. This merely reflects that
in the large-$N$ limit, we have already incorporated {\em all}
quantum fluctuations of $\phi_1 \cdots \phi_N$.

The potential $V(\phi^2/N)$ that leads to this effective action
is not physically well-defined. The same is true about the
quantum fluctuations coming from the path integral measure.
Unlike the effective action, neither $V$ nor the measure is
finite when regulators are removed. Furthermore, neither the
quantum fluctuations nor $V$ is scale invariant. They depend on
a scale $M$, which however mutually cancels to give a finite
and scale free effective action. $V(\phi^2/N)$ is not in any
sense an approximation to the effective potential i.e., $\G$
for constant backgrounds. Nevertheless, $V$ does appear in
intermediate stages of calculations and many people want to
know what it is. Its `finite part' in zeta function
regularization in the large-$N$ limit, grows as
$V(\fr{|\phi|^2}{N}) \sim \fr{|\phi|^4/N^2}{\log(\bar \la
|\phi|^2/M^2 N)}$ for large $|\phi|^2/N$ (appendix
\ref{a-back-to-V-legendre-transform}). We have not yet determined its
behavior for small $|\phi|^2/N$, though there is no indication of singular behavior.
$V(\eta)$ is most easily
expressed in terms of the Laplace transformed potential
$W(\si)$. At $N=\infty$ the finite part of $W(\si)$ in zeta
function regularization is ($e$ is the base of natural
logarithms)
    \beq
        W_0(\si) = \la \si^2 - \fr{\hbar}{16\pi^2}
        \bigg\{ \half \si_o^2 \log[{\si_o e^{-3/2} \over M^2}] + \si_o
        \log[{\si_o \over M^2 e}] (\si - \si_o) + \half \log[{\si_o \over M^2}] (\si - \si_o)^2
        \bigg\}.
    \eeq
$\si_o$ is the constant background value of the field $\si(x)$
appearing in the effective action. Though $\G_0$ involves
arbitrarily high derivatives, $W_0$ does not, indicating a form of
locality. It is interesting to know whether this remains true in other regularization
schemes and after including $1/N$ corrections.

Next, we deform this line of fixed points by adding a mass term
$-m^2 \si$ to $W(\si)$, which explicitly breaks
scale-invariance. In section
\ref{s-const-extrema-of-eff-action} we determine constant
extrema of the large-$N$ effective action and find a phase $b =
0, \si = {m^2 \over 2\la}$ where $O(N+1)$ is unbroken and
another phase $b = \pm m, \si=0$, where $O(N+1)$ symmetry is
spontaneously broken to $O(N)$. In section
\ref{s-mass-of-lightest-particles} we determine the masses of
the lightest scalar excitations in the $O(N+1)$ symmetric phase
    \beq
    M_b = {m \over \sqrt{2\la}} ~~~~~~{\rm and} ~~~~ M_\si = {4 \sqrt{3} \pi m
    \over  \sqrt{\hbar}}.
    \eeq
We are not yet able to study the broken phase; it occurs at $\si =0$.
Though $W(\si)$ is singular at $\si =0$, the effective action $\G$
may be regular there (the effective potential is), we do not know.

In section \ref{s-leading-1-ov-N-corrns} we attempt to preserve
scale-invariance after including quantum fluctuations in $\si$ and
$b$ at order $1/N$. Some quantum effects were already present at
$N=\infty$, due to fluctuations of $\phi_1, \cdots, \phi_N$ which we
integrated out. Determining the effective action at order $1/N$ is
complicated, so we make some further approximations
which we hope to relax later. In sec. \ref{s-gamma-1-for-B-equal-0}
we assume that quantum fluctuations are slowly varying and determine
the $1/N$ correction to the fixed point potential, $W_0 + \N W_1$
that ensures the effective action $\G(B=0,\Si) = \G_0 + \N \G_1$ is
finite \& scale-invariant for general background fields $\Si(x)$.
But there could be further divergences not canceled by $W_1(\Si)$,
since it is independent of $B$. In sec. \ref{s-RGE-for-gamma-1} we
drop the assumption $B=0$, but assume that background fields are
constant and derive a Callan-Symanzik renormalization group equation
(RGE) for the effective action in zeta function regularization.
Imposing the condition that the $\gb$ function of $\la$ vanishes and
using the previous result for $W_1(\Si)$ leads to a unique solution
for the $\gb$-function of the mass and anomalous dimensions of $b$
and $\si$ fields. Within our (drastic) approximations, the model is
renormalizable and the line of fixed points is maintained at order
$1/N$. It remains to analyze the phase in which $O(N+1)$
symmetry is broken, and also couple our model to fermions and gauge fields.
Finally, we wonder whether there might be a dual/holographic/string
description of our scale-invariant model by analogy
with the AdS/CFT conjecture \cite{string-dual}.

\section{Lagrangian and change of field variables}
\label{s-lagrangian-change-of-var}

Consider an $N+1$ component real scalar field $\phi_i, i=
0,2,\cdots, N$ in $4$ Euclidean space-time dimensions. We require
the action to be globally $O(N+1)$ invariant. The partition function
is
    \beq
    Z = \int [D\phi] e^{-\ov{\hbar} \int d^4x \half
        \{|\grad \phi_i|^2 + N V(\phi^2/N) \}}.
    \label{e-original-partition-function}
    \eeq
The factors of $N$ have been chosen to facilitate a meaningful large-$N$ limit.
$\phi^2$ is short for $\sum_{i=0}^{N} \phi_i \phi_i$.
$V(\fr{\phi^2}{N})$ will be determined by requiring the quantum
effective action to be scale-invariant. We will find the effective
action in an expansion around $N=\infty$ holding $\hbar$ fixed. In
this limit, $\eta(x) = \phi^2/ N \geq 0$ has small fluctuations and behaves
classically but, $\phi_i$ continue to have large fluctuations. We
will integrate out all the $\phi_i$ except $\phi_0$ and study the
model in an expansion around $N=\infty$, where $\eta$ and $b =
\phi_0/\sqrt{N}$ have small fluctuations. We impose $\eta =
\phi^2/N$ via a delta function
    \beq
        Z = \int [D\phi] \int_0^\infty [D\eta] e^{-\ov{\hbar} \int d^4x \half
        \{ |\grad \phi|^2 + NV(\eta) \} } \prod_x \delta(N \eta(x) -
        \phi^2(x)).
    \eeq
Now insert the integral representation
    \beq
    \delta(N\eta - \phi^2) = \int_{\cal C} \fr{d\si}{2\pi
        i} e^{\si(N\eta - \phi^2)}.
    \eeq
$\cal C$ can be any contour from $-i\infty$ to $i\infty$ since
the integrand is entire.
Up to an overall constant that cancels from normalized correlations,
(\ref{e-original-partition-function}) becomes
    \beq
    Z = \int [D\phi] \int_0^\infty [D\eta]
        \int_{\cal C} [D\si] e^{-\ov{\hbar} \int d^4x \half
        \{ |\grad \phi|^2 + \si \phi^2 + N V(\eta) - N \si \eta \}
        }.
    \eeq
$\si$ is a Lagrange multiplier enforcing the
constraint $\eta = \phi^2/N$. Though $\cal C$ is not the real line,
$Z$ is real by construction. We separate
$b = \phi_0/\sqrt{N}$ in anticipation of integrating out $\phi_1, \ldots, \phi_N$,
    \beq
    Z = \int [Db] \int [D\phi] \int_0^\infty [D\eta] \int_{\cal C} [D\si]
        e^{-\ov{2\hbar} \int d^4x
        \bigg[ \sum_{i=1}^N \{(\grad \phi_i)^2 + \si \phi_i^2 \}
        + N (\grad b)^2 + N \si b^2 + N V(\eta) - N \si \eta
        \bigg]}.
    \eeq
Henceforth $[D\phi]$ does not include $\phi_0$. Now, we reverse
the order of $\eta$ and $\si$ integrals and observe that the
$\eta$ integral is a Laplace transform at each space-time point
$x$,
    \beq
    \int_0^\infty [D\eta] e^{-(N / 2\hbar) \int d^4x [ V(\eta) - \si \eta]}
        = e^{-({N / 2\hbar}) \int d^4x W(\si)}.
    \label{e-V-to-W-laplace transform}
    \eeq
Then
    \beq
    Z = \int [Db] \int [D\phi] \int_{\cal C} [D\si]
    e^{-(1/2\hbar) \int d^4x \bigg[ \sum_{i=1}^N ((\grad \phi_i)^2 + \si
    \phi_i^2) + N (\grad b)^2 + N \si b^2 + N W(\si)  \bigg]}.
    \eeq
Reversal of $\si$ and $\eta$ integrals is allowed if $W(\si)$
is non-singular along contour $\cal C$. Now we reverse the order of the $\phi$ and $\si$
integrals. The $\phi$ integral is a gaussian
    \beq
    \int [D \phi] e^{-(1/2\hbar) \sum_{i=1}^N \int d^4x \phi_i (-\grad^2+ \si)
    \phi_i}  = \bigg[ \det{\bigg( \fr{-\grad^2 + \si}{2\pi \hbar}\bigg)}
    \bigg]^{-N/2}.
    \eeq
This integral converges if $-\grad^2 + \si$ has eigenvalues with
positive real part. Since $-\grad^2$ is a positive operator, this is
ensured if $\Re \si > 0$ (though the answer possesses an analytic
continuation even to $\si$ with negative real part, as long as it
stays off the negative real axis). Thus we get (up to an irrelevant
overall constant)
    \beq
    Z = \int_{-\infty}^\infty [Db] \int_{\cal C} [D\si] e^{-N S(b,\si)}
    \eeq
where
    \beq
    S(b,\si) = \ov{2\hbar} \bigg[\hbar
    \tr \log (-\grad^2 + \si) + \int d^4x \bigg\{ (\grad b)^2
        + \si b^2 + W(\si) \bigg\} \bigg].
    \label{e-action}
    \eeq
Here, $W(\si)$ is obtained from $V(\eta)$ via a Laplace
transform (\ref{e-V-to-W-laplace transform}), at each $x$
    \beq
    e^{-(N/2\hbar)W(\si(x))} = \int_0^\infty e^{-(N/2\hbar) (V(\eta(x)) -
    \si(x) \eta(x))} d\eta(x).
    \eeq
Conversely, $V(\eta)$ is obtained from $W(\si)$ by an inverse
Laplace transform along contour ${\cal C}$ to the right of all
singularities of $W(\si)$
    \beq
    \int_{\cal C} \fr{d\si}{2\pi i} e^{-(N/2 \hbar) (W(\si) + \si
    \eta)} = e^{-(N/2\hbar)V(\eta)}.
    \eeq
$\si$ is a dynamical field, it carries space-time derivatives
and complicated self interactions. We are interested in
correlation functions of the $\si$ and $b$ fields, which are
defined as
    \beq
        \langle \si(x_1) \cdots \si(x_n) b(y_1) \cdots b(y_m) \rangle =
        \ov{Z} {\int [Db] \int_{\cal C} [D\si]  e^{-N S(b,\si)}
        \si(x_1) \cdots \si(x_n) b(y_1) \cdots b(y_m)}.
    \label{e-def-of-corrlns-b-and-sig}
    \eeq
Positivity of $\eta = \phi^2/N$ implies that $\si(x)$ is valued on a
contour $\cal C$ from the south pole to north pole of the complex
plane. On the other hand, the contour of integration for the $b$
field is the real line. $\si$ has dimensions of mass$^2$ while $b$
has dimensions of mass.

\section{Scale-invariance of the effective action at $N=\infty$}
\label{s-scale-inv-of-effac-inf-N}

The interaction $W(\si)$ appearing in the action
(\ref{e-action}) is expanded in inverse powers of $N$
    \beq
        W(\si) = W_0(\si) + \N W_1(\si) + \ov{N^2}
            W_2(\si) + \cdots.
    \eeq
We do not assume analyticity of $W(\si)$ at $\si = 0$, so we do not
expand it in powers of $\si$. $W(\si)$ is to
be determined by the principle that the theory be scale-invariant
at each order in $1/N$. Of course, the action $S(b,\si)$ is also
expanded in inverse powers of $N$
    \beq
        S(b,\si) &=& \ov{\hbar} \bigg[S_0 + \N S_1 + \ov{N^2} S_2
            + \cdots \bigg] \cr
    {\rm where~~~}
        S_0 &=& \half \bigg[\hbar \tr \log[-\grad^2 + \si] +
            \int d^4x \bigg\{(\grad b)^2 + \si b^2 + W_0(\si) \bigg\}
            \bigg], \cr
        S_1 &=& \half \int d^4x W_1(\si), ~~~
        S_2 = \half \int d^4x W_2(\si) ~~~~ {\rm etc}.
    \eeq
$W_{1,2,3\cdots}$ must be chosen to cancel divergences and
scale anomalies coming from fluctuations in $b$ and $\si$ while
$W_0$ is chosen to cancel those from fluctuations in $\phi_1
\cdots \phi_N$. The possible choice(s) of $W_{0,1,2\ldots}$ define the scale invariant fixed
point(s) just as $\half |\pdr \phi|^2$ defines the trivial fixed point. $W_n$ are not counter terms in the perturbative sense. For a given choice of $W_n$ that does the job, there may be interactions with arbitrary coupling constants we can add to $W_n$ and preserve scale invariance. We want this to be a finite parameter family. $W_n$ are not restricted to be of any particular form. For locality, we wish to avoid arbitrarily high derivatives of $\si$.

$N$ and $\hbar$ appear differently in $S(b,\si)$. $N \to
\infty$ is a `classical limit' in which $b,\si$ have
small fluctuations and is governed by the action
$S_0(b,\si)$. $\hbar \to 0$ is also a classical limit, one in
which the original fields $\phi_i$ have small fluctuations. It
is governed by the original action $\int d^4x [|\grad \phi|^2 +
N V(\phi^2/N)]$. These two classical limits potentially capture
different features of the full quantum theory for a given
$W(\si)$. There is {\em a priori} no reason for the two limits
$\hbar \to 0$ and $N \to \infty$ to commute. $\hbar
\tr\log[-\grad^2 + \si]$ is a quantum correction
to the action in the $\hbar \to 0$ limit. But it is part of the `classical' action in the $N \to \infty$
limit.

A theory is scale-invariant if its effective action $\Gamma$ (Legendre transform of the
generating series of connected correlations,
which generates all 1PI or proper vertices, see appendix
\ref{a-eff-action}), is scale-invariant.
Such an effective action defines a fixed point of the
renormalization group flow. $\Gamma$ is obtained by averaging
over fluctuations in $b$ and $\si$ and is defined implicitly
by
    \beq
    e^{- N \Gamma(\B,\Si)} = \int [D\beta] \int_{\cal C}
        [D\vsi] \exp\bigg[-N \bigg\{S(\B+\beta,\Si+\vsi) - \beta \fr{\delta \Gamma}{\delta \B}
         - \vsi \fr{\delta \Gamma}{\delta \Si} \bigg\} \bigg].
    \label{e-implicit-def-of-eff-action}
    \eeq
$\B(x)$ and $\Si(x)$ are arbitrary background fields while $\beta$
and $\vsi$(`varsigma') are the fluctuating fields,
$b = \B + \beta, ~~ \si = \Si + \vsi$.
$\Gamma(\B,\Si)$ is calculated in a series in powers of $1/N$
holding $\hbar$ fixed
    \beq
        \Gamma(\B,\Si) =  \Gamma_0(\B,\Si) + \N \Gamma_1(\B,\Si) +
        \ov{N^2} \Gamma_2(\B,\Si) + \cdots.
    \eeq
To zeroth order in $1/N$, the effective action
    \beq
        \Gamma_0(\B,\Si) = S_0(\B,\Si) = \half \bigg[
            \hbar \tr \log[-\grad^2 + \Si]
            + \int d^4x \bigg\{(\grad \B)^2
            + \Si \B^2 + W_0(\Si) \bigg\}
        \bigg].
    \label{e-def-of-eff-ac-at-infinite-N}
    \eeq
${\rm Tr} \log[-\grad^2 + \Si(x)]$ is
divergent. We must regulate the model so
that it is finite. Then we must pick
$W_0(\Si)$ (which will depend on the regulator) in such a way
that when the regulator is removed, $\Gamma_0(B,\Si)$ is not
just finite but also scale-invariant. Similarly, $W_1$ is
determined by the principle that $\Gamma_1$ be scale-invariant
and so on. Actually, we will also have
to allow for wave function renormalizations, but these appear
only at order $1/N$, see sec. \ref{s-leading-1-ov-N-corrns}.

\subsection{Effective action for constant $\si$ at $N = \infty$}
\label{s-cl-eff-ac-const-bkgrnd}

${\rm Tr~} \log[-\grad^2 + \Si(x)]$ appearing in the large-$N$
effective action $\Gamma_0(B,\Si)$
(\ref{e-def-of-eff-ac-at-infinite-N}) is most easily evaluated
for a constant background $\Si(x) = \Si_o$. This is a
physically reasonable first approximation if space-time
inhomogeneities are small. $\Si(x)$ takes values on the
contour $\cal C$ from $-i\infty$ to $i \infty$. We
anticipate needing to pick the contour to avoid the negative
real axis, so $\Si_o$ will be a complex
number that lies off the negative real axis.

\subsubsection{Momentum cutoff regularization}
\label{s-mom-cut-off-regularization}

In momentum cutoff regularization,
    \beq
    \tr \log[-\grad^2 + \Si_o] &=& \int_{|p|<\La} d^4p~ d^4q~ \tl \gd(p-q) \tl
        \gd(p-q) \log(p^2 + \Si_o) = \tl \gd(0) \int_{|p|< \La} d^4 p \log[p^2 +
        \Si_o] \cr
    &=& {\Om \over (2\pi)^4} \int_0^\La dp~ p^3 ~ \log[p^2 + \Si_o]
    \int d\Om_4.
    \eeq
Here the space-time volume is $\int d^4x = \Om = (2\pi)^4 \tl
\gd(0)$ and $\int d\Om_4 = 2\pi^2$ is the `surface area' of a
unit $3$-sphere embedded in 4d Euclidean space.
Thus,
    \beq
    \hbar \tr \log[-\grad^2 + \Si_o] &=& {\hbar \Om \over 8 \pi^2}
        \int_0^\La dp~ p^3 \log(p^2 + \Si_o) \cr
    &=& {\hbar \Om \over 64 \pi^2} \bigg[2 \La^4 \log(\La^2 + \Si_o) - \La^4
        + 2 \La^2 \Si_o - 2 \Si_o^2 \log(\La^2 + \Si_o) + 2 \Si_o^2 \log\Si_o
        \bigg] \cr
    &=& {\hbar \Om \over 64 \pi^2} \bigg[ 2 \La^4 \log \La^2 -
        \La^4 + 4 \La^2 \Si_o - 2 \Si_o^2 \log \La^2 ~~({\rm ~divergent~ terms}) \cr
    && + 2 \Si_o^2 \log \Si_o ~~({\rm non~scale~invariant~finite~term}) \cr
    && - \Si_o^2 ~~({\rm scale~invariant~ finite~ term})
    + {\rm ~terms~ that~ vanish~ as~} \La \to \infty \bigg].
    \eeq
We must pick $W_0(\Si)$ such that the large-$N$ effective
action (\ref{e-def-of-eff-ac-at-infinite-N}) is both finite and
scale-invariant when $\La \to \infty$. In sec.\ref{s-cl-eff-ac-arbit-bkgrnd}
we do this for general $\Si,\B$. Here we get an idea of the answer by requiring that
$\Gamma_0(\B,\Si)$ be scale-invariant for constant $\Si =
\Si_o$. To this end, we can pick
    \beq
    W_0(\Si_o,\La)  = {- \hbar \over 64 \pi^2} \bigg[2 \La^4 \log \La^2
        - \La^4 + 4 \La^2 \Si_o - 2 \Si_o^2 \log \La^2 \bigg]
        - {\hbar \over 32 \pi^2} \Si_o^2 \log \Si_o.
    \eeq
This is the `minimal subtraction' choice. Our principle that
$\G_0(B,\Si_o)$ be finite and scale-free is ambiguous. We could
add to this choice of $W_0(\Si_o)$, any scale-free finite term
of the form $\la \Si_o^2$, where $\la$ is a dimensionless
coupling constant. Other terms such as $m^2 \Si_o$ or terms
proportional to a higher power of $\Si_o$ would involve
dimensional coupling constants and would explicitly introduce a
scale into the theory. The general choice leading to a
scale-free $\G_0(B,\Si_o)$ is
    \beq
    W_0(\Si_o,\La)  = {- \hbar \over 64 \pi^2} \bigg[2 \La^4 \log \La^2
        - \La^4 + 4 \La^2 \Si_o - 2 \Si_o^2 \log \La^2 \bigg]
        - {\hbar \over 32 \pi^2} \Si_o^2 \log \Si_o + \la \Si_o^2.
    \eeq
In the large-$N$ limit, we have a $1$-parameter family of RG fixed
points, parameterized by $\la$. The addition of $m^2
\Si_o$ corresponds to a relevant perturbation of one of
these fixed points. The addition of $c_n \Si_o^n$ for $n
> 2$ corresponds to an irrelevant deformation, since coupling $c_n$ has a
negative mass dimension. In the sequel we consider the
mass deformed theory, where $W_0(\Si_o)$ is a two-parameter
$(m,\la)$ family
    \beq
    W_0(\Si_o,\La)  = {- \hbar \over 64 \pi^2} \bigg[2 \La^4 \log \La^2
        - \La^4 + 4 \La^2 \Si_o - 2 \Si_o^2 \log \La^2 \bigg]
        - {\hbar \over 32 \pi^2} \Si_o^2 \log \Si_o + \la
        \Si_o^2 - m^2 \Si_o.
    \eeq
$W_0(\Si_o)$ has a branch cut along the negative $\Si_o$
axis, consistent with our expectation that $\Si(x)$ is valued
on a contour that misses the negative real axis.
The corresponding large-$N$ effective action
(\ref{e-def-of-eff-ac-at-infinite-N}) for constant backgrounds
$B_o,\Si_o$ is
    \beq
    \G_0(B_o,\Si_o) = {\Om \over 2} \bigg[ -m^2 \Si_o + \bigg(\la - {\hbar \over 64 \pi^2}
        \bigg) \Si_o^2 + \Si_o B^2 \bigg].
    \eeq
Having found a line of fixed points, are they UV or IR? The
answer can depend on the direction in which we flow from a fixed point.
In this case, all the above fixed points are UV with
respect to the mass deformation. The physical reason is that
as we go to higher energies, the ratio of $m$ to
the energy scale will decrease and the RG
flow will tend towards the fixed point. Recall that the
gaussian fixed point (massless free scalar field theory) in $4$d
is UV with respect to mass deformations, but IR with respect to the quartic coupling $\la
\phi^4$. So far, in our model, the analogue of the quartic
coupling, $\la \Si_o^2$ is exactly marginal. Note that $\Si$
has a canonical dimension of mass squared, while $\phi$ has
dimensions of mass.

The presence of a $1$-parameter family of UV fixed points is a godsend.
It means that for {\em any}
values of $\la$ and $\hbar$, we can set $m=0$ and get an extra
symmetry, scale-invariance. Thus, the
mass parameter can be naturally small (at least at $N=\infty$).
An interesting question is whether the line
of fixed points can be maintained after including effects of
quantum fluctuations in $\si$ and $b$ in a $1/N$ expansion (see sec. \ref{s-leading-1-ov-N-corrns}).

\subsubsection{Zeta function regularization}
\label{s-zeta-fn-regularization}

We repeat the evaluation of the large-$N$ effective potential
by regularizing $\tr \log[-\grad^2 + \Si_o]$ via zeta-function
regularization. This method directly prescribes a finite part
for $\tr \log[-\grad^2 + \Si_o]$ in the unregulated limit. The
finite part is not scale-invariant. We will pick $W_0(\Si_o)$ to
cancel this non-scale-invariant quantity, so that
$\Gamma_0(B,\Si_o)$ is both scale-invariant and finite. This
procedure means we do not need to prescribe how $W_0(\Si_o)$ must
depend on the regulator, but just the finite part of its limiting unregulated
value. Such a short-cut is not possible in other regularization
schemes such as momentum cutoff (sec.
\ref{s-mom-cut-off-regularization}) or dimensional
regularization (appendix \ref{a-dim-reg}). For this and other reasons, we
will use the zeta-function regularization in the rest of
the paper. On the other hand, comparison of different schemes
allows us to better understand what is scheme independent. We
find that the presence of a $1$-parameter family of fixed
points is scheme independent. Moreover, the finite and
scale-invariant effective potential obtained in the three schemes
are the same up to a finite relabeling of $\la$.
To define the finite part of $\tr \log[-\grad^2 + \Si_o]$ by
zeta-function regularization, let
    \beq
    \zeta(s) = \tr [-\grad^2 + \Si_o]^{-s}
    = \int d^4p~ d^4q~ \tl\gd^4(p-q) {\tl\gd^4(p-q) \over [p^2 + \Si_o]^s}
    = \Om \int \fr{d^4p}{(2\pi)^4}~ \ov{[p^2 + \Si_o]^s}.
    \eeq
$\Om = \int d^4x = (2\pi)^4 \tl \gd^4(0)$,
appears because $\Si(x)$ is homogeneous and we are taking the
trace of an operator that is diagonal in momentum space.
$\zeta(s)$ is analytic for $\Re s > 2$. By analytic continuation, we get a meromorphic function
$\zeta(s)$. Away from singularities,
    \beq
        \zeta^\prime(s) = -\tr \bigg[\fr{\log[-\grad^2 + \Si_o]}{[-\grad^2 +
        \Si_o]^s} \bigg].
    \eeq
So, if $\zeta(s)$ is analytic at $s=0$,  $\zeta^{\prime}(0)= -\tr \log[-\grad^2 + \Si_o]$.
Now let us calculate
    \beq
        \fr{\zeta(s)}{\Om} =
            = \ov{(2\pi)^4}\int d\Om_4 \int_0^\infty \fr{p^3 dp}{(p^2 +
            \Si_o)^s}
        = \ov{(2\pi)^4} 2\pi^2 \fr{\Si_o^{2-s}}{2(s-1)(s-2)}
        = \ov{16 \pi^2} \fr{\Si_o^{2-s}}{(s-1)(s-2)}.
    \label{e-zeta-of-s-const-bkgrnd}
    \eeq
$\zeta(s)$ has simple poles at $s = 1,2$ but is analytic at
$s=0$ so,
    \beq
    \fr{\zeta^{\prime}(0)}{\Om} = \fr{3\Si_o^2}{64 \pi^2} -
            \fr{\Si_o^2 \log \Si_o}{32\pi^2}
    &=& - \fr{\Si_o^2 \log[\Si_o e^{-3/2}]}{32 \pi^2} \cr
        \Rightarrow ~~~ \tr \log[-\grad^2 + \Si_o]
    &=& \fr{\Si_o^2 \Om}{32 \pi^2} \log[{e^{-3/2} \Si_o}].
    \label{e-tr-log-const-bkgrnd}
    \eeq
(\ref{e-tr-log-const-bkgrnd}) is not scale-free due to the
logarithm. The effective action at $N=\infty$ for constant
background fields $B,\Si$ is
    \beq
        \Gamma_0(B_o,\Si_o) = \fr{\Om}{2} \bigg[ \fr{\hbar \Si_o^2}{32 \pi^2}
        \log[e^{-3/2} {\Si_o \over M^2}] + \Si_o B_o^2 + W_0(\Si_o) \bigg].
    \label{e-eff-ac-at-inf-N-const-bkgrnd}
    \eeq
The parameter $M$ with dimensions of mass sets the scale for the logarithm
and breaks scale-invariance. The choice of $W_0(\si)$ that
ensures $\Gamma_0(\B,\Si_o)$ is scale-free (for $m=0$) is
    \beq
    W_0(\si_o) = -m^2 \si_o + \la \si_o^2 - \fr{\hbar \si_o^2}{32 \pi^2}
        \log[e^{-3/2} {\si_o \over M^2}].
    \label{e-W_0-for-constant-bkgrnd}
    \eeq
We added a mass term, a relevant perturbation away from the line of
fixed points parameterized by $\la$. We will prove in
sec. \ref{s-cancel-scale-anomaly} that the scale anomaly in
$\Gamma_0(\B_o ,\Si_o)$ vanishes for this choice of
$W_0(\si_o)$. We avoid cubic and higher powers of $\si_o$ as
before. Note that the terms in $W_0(\si_o)$ are of different
orders in $\hbar$ but of the same order in $\ov{N}$. For this
choice of $W_0$, we get
    \beq
    \Gamma_0(\B_o,\Si_o) = \fr{\Om}{2} [-m^2 \Si_o
    + \la \Si_o^2 + \Si_o \B_o^2 ].
    \label{e-gamma0-zeta-fn-const-bkgrnd}
    \eeq
$M$ cancels out from the
effective potential, which is scale-free for $m=0$. Though
$W_0(\si_o)$ has a branch cut along the negative
$\Si_o$ axis, $\Gamma_0(\B_o,\Si_o)$ (and
$S_0(b_o,\si_o)$) is entire. It is very interesting
to know whether this is true for general backgrounds. The method of expanding
in inverse powers of $\Si_o$ that we use in the rest of this paper prevents
us from answering this question here. Comparing with
section \ref{s-mom-cut-off-regularization} we see that
irrespective of regularization scheme, there is a one parameter
family of fixed points parameterized by $\la$. However, the
definition of the coupling depends on regularization scheme,
    \beq
    \la_{\rm zeta~function} = \la_{\rm cutoff} - {\hbar \over 64 \pi^2} .
    \eeq

\subsection{$N = \infty$ effective action expanded around a constant background}
\label{s-cl-eff-ac-arbit-bkgrnd}

In section \ref{s-cl-eff-ac-const-bkgrnd} we calculated the
$N=\infty$ effective action
(\ref{e-def-of-eff-ac-at-infinite-N}) for constant $\B,\Si$.
Allowing for arbitrary backgrounds $B(x)$ is easy, the difficulties
lie in non-constant $\Si(x)$. Here, we get an expansion for
$\Gamma_0(\B,\Si)$ in powers and derivatives of $\Si - \Si_o$
where $\Si_o$ is a constant background whose value is not a
negative real number or zero. It would be interesting to
calculate $\G_0$ by complementary methods too.
From appendix \ref{a-exp-for-tr-log}, in zeta
function regularization,
    \beq
    \tr\log[-\grad^2 + \Si(x)] &=& {\Si_o^2 \Om \over 32\pi^2}
        \log[\Si_o e^{-3/2}] + \int {d^4x \over 16\pi^2} \bigg[
        \Si_o \log[{\Si_o/e}] (\Si - \Si_o) + \half (\Si -
        \Si_o)^2 \log{\Si_o}
            \cr && - (\Si - \Si_o) \Pi(\Delta) (\Si - \Si_o) + {\cal O}(\Si - \Si_o)^3 \bigg].
    \label{e-tr-log-arbit-bkgrnd}
    \eeq
This reduces to (\ref{e-tr-log-const-bkgrnd}) for constant
backgrounds ($\Si= \Si_o$). The ${\cal O}(\Si - \Si_o)^3$ and
higher order terms can be obtained by the method of appendix
\ref{a-exp-for-tr-log}. But these higher order terms are
scale-invariant. We will show in section
\ref{s-cancel-scale-anomaly} that the part of $\tr
\log[-\grad^2 + \Si(x)]$ that is not scale-invariant, is
restricted to the first three terms on the rhs of
(\ref{e-tr-log-arbit-bkgrnd}). Note that $\Delta = -\grad^2
/\Si_o$ and
    \beq
        \Pi(\Delta) = \sum_{n=1}^\infty {(-\Delta)^{n}
            \over n(n+1)(n+2)} = {\Delta(3 \Delta +2) -2(\Delta+1)^2
            \log{(1+\Delta)} \over 4 \Delta^2}
        = -{\Delta \over 6} + {\Delta^2 \over 24} - {\Delta^3
            \over 60} + \cdots
    \eeq
Thus, the effective action at $N=\infty$ is
    \beq
        \Gamma_0(\B,\Si) &=& \half \int d^4x \bigg[(\grad \B)^2 + \si \B^2 + W_0(\Si) +
        \fr{\hbar}{16\pi^2}
        \bigg\{\half \Si_o^2 \log[\Si_o e^{-3/2}] + \Si_o
        \log[\Si_o/e] (\Si - \Si_o) \cr && + \half \log[\Si_o] (\Si - \Si_o)^2
        - (\Si - \Si_o) \Pi(\Delta) (\Si - \Si_o) + {\cal O}(\Si - \Si_o)^3
        \bigg\}
        \bigg].
    \label{e-eff-ac-large-N-arbit-bkgrnd}
    \eeq
In deriving (\ref{e-eff-ac-large-N-arbit-bkgrnd}) we did {\em
not} assume $\Si$ is slowly varying but rather that $\Si -
\Si_o$ is small and that $\Si_o$ is not a negative real number or zero.

\subsection{Fixing the interaction at $N=\infty$ by requiring scale-invariance}
\label{s-fix-W_0-arbit-bkgrnd}

$W_0(\si)$ must be chosen so that $\Gamma_0(\B,\Si)$ in
(\ref{e-eff-ac-large-N-arbit-bkgrnd}) is scale-free. The choice
that does the job is
    \beq
        W_0(\si) = \la \si^2 -m^2 \si - \fr{\hbar}{16\pi^2}
        \bigg\{ {\si_o^2 \over 2} \log[{\si_o e^{-{3 \over 2}} \over M^2} ] + \si_o
        \log[{\si_o \over M^2 e}] (\si - \si_o) + \half \log[{\si_o \over M^2}] (\si - \si_o)^2 \bigg\}
    \label{e-W_0}
    \eeq
with $m=0$. For $m \ne 0$ we have a mass deformation.
$M$ sets the scale for logarithms, but cancels
out in the $N=\infty$ effective action
    \beq
        \Gamma_0 = \int {d^4x \over 2} \bigg[
            (\grad \B)^2 + \Si \B^2 - m^2 \Si + \la \Si^2 - \fr{\hbar}{16\pi^2}
            \bigg\{(\Si - \Si_o) \Pi(\Delta) (\Si - \Si_o)
         + {\cal O}(\Si - \Si_o)^3 \bigg\} \bigg].
    \label{e-eff-ac-large-N-finite-and-scale-invariant}
    \eeq
$\Gamma_0(B,\Si)$ is now free of divergences (for $\Si_o \notin {\bf R}^-$) and defines
the large-$N$ effective action for background fields $\Si$
whose deviation from a constant $\Si_o$ is small. The higher
order terms in $\vsi = \Si - \Si_o$ are all finite, scale-free
and calculable by the method of appendix
\ref{a-exp-for-tr-log}. The fact that $\Gamma_0(B,\Si)$ is not quadratic
when regulators are removed, indicates that our theory is not
trivial.

\subsection{Cancelation of scale anomaly}
\label{s-cancel-scale-anomaly}

We show that the effective action $\G_0(B,\Si)$ in
(\ref{e-eff-ac-large-N-finite-and-scale-invariant}) is scale
invariant for $m=0$ ($m \ne 0$ is treated subsequently).
Coordinates and fields are rescaled according to
their canonical dimensions. $M$ sets the scale for logarithms
and explicitly introduces a scale into the theory. To encode
this physical fact, $M$ is {\em not} rescaled, the same is true
of $m$. We define dilations as
    \beq
        D_a x^\mu = a^{-1} x^\mu, ~~ D_a b = a b, ~~ D_a \si = a^2 \si,
        ~~ D_a \la = \la, ~~ D_a \grad = a \grad.
    \label{e-scale-transformation}
    \eeq
The generator of infinitesimal dilations $\delta_D$ is defined
as
    \beq
        \delta_D f &=& \lim_{\eps \to 0} {D_{1+\eps} f - f \over
        \eps},
            \cr
        \delta_D x^\mu = - x^\mu, ~~ \delta_D b = b, ~~ \delta_D \si = 2 \si,
        && \delta_D \la = 0,  ~~ \delta_D^0 \grad = \grad, ~~
        \delta_D dx = -dx
    \label{e-dilatation-generator}
    \eeq
and may be represented as
    \beq
        \delta_D = -x^\mu \dd{}{x^\mu} + b(x) \dd{}{b(x)} + 2 \si(x) \dd{}{\si(x)}.
    \eeq
We now show that $\Gamma_0(B,\Si)$
(\ref{e-def-of-eff-ac-at-infinite-N}) is invariant under
dilations for (not necessarily constant) backgrounds $B(x)$ and
$\Si(x)$ if $W_0(\Si)$ is chosen according to (\ref{e-W_0})
with $m=0$. First,
    \beq
    D_a \int d^4x \{ (\grad B)^2 + \Si B^2 \} = \int d^4x  \{ (\grad B)^2 + \Si B^2
        \} ~~\Rightarrow  ~~ \delta_D \int d^4x \{ (\grad B)^2 + \Si B^2 \} =
        0.
    \eeq
$W_0(\Si)$ and $\tr \log[-\grad^2 + \Si]$ are the only terms in
(\ref{e-def-of-eff-ac-at-infinite-N}) with non-trivial (in fact
inhomogeneous) scale transformations. $W_0$ transforms as
    \beq
    D_a \int d^4x W_0(\Si) &=& \int d^4x W_0(\Si) - \fr{\hbar}{16\pi^2} \int d^4x
        \bigg[ \half \Si_o^2 \log[a^2]
        + \Si_o \log[a^2] \vsi + \half \log[a^2] \vsi^2
        \bigg] \cr
    &=& \int d^4x W_0(\Si) - \fr{\hbar \Si_o^2 \Om \log a}{8\pi^2}
        \bigg[\half + \fr{\bra \vsi \ket}{\Si_o} + \fr{\bra \vsi^2 \ket}{2 \Si_o^2} \bigg]
    \cr
    \Rightarrow ~~ \delta_D \int d^4x W_0(\Si) &=& - \fr{\hbar \Om \Si_o^2}{8
        \pi^2} \bigg[\half + \fr{\bra \vsi \ket}{\Si_o} + \fr{\bra \vsi^2 \ket}{2 \Si_o^2} \bigg]
    \eeq
where $\vsi = \Si - \Si_o$, $\Si_o$ is a constant background and
$\bra f \ket = \ov{\Om} \int d^4x f $. On the other hand,
    \beq
        \tr \log[-\grad^2 + \Si] = - \zeta^{\prime}(0)
        &{\rm and}& \zeta(s) = \tr [-\grad^2 + \Si]^{-s} \cr
    \Rightarrow ~~~~ D_a \zeta(s) = a^{-2s} \zeta(s)
        & {\rm and}& D_a \zeta^{\prime}(s) = - 2 \zeta(s) a^{-2s}\log a
         + a^{-2s} \zeta^{\prime}(s).
    \eeq
Now set $s=0$ and use the result for $\zeta(0)$ calculated in
Appendix \ref{a-scale-anomaly},
    \beq
        D_a \zeta^{\prime}(0) = \zeta^{\prime}(0) - 2 \zeta(0) \log a
        &\implies& \delta_D \zeta^\prime(0) = -2 \zeta(0)
            =  - {\Om \Si_o^2 \over 8 \pi^2} \bigg[
            \half + {\bra \vsi \ket \over \Si_o} + {\bra \vsi^2 \ket \over 2 \Si_o^2}
            \bigg], \cr
    \implies ~~  \delta_D \hbar \tr \log[-\grad^2 + \Si] &=& - \hbar \delta_D
        \zeta^\prime(0) = {\hbar \Om \Si_o^2 \over 8 \pi^2} \bigg[
            \half + {\bra \vsi \ket \over \Si_o} + {\bra \vsi^2 \ket \over 2 \Si_o^2}
            \bigg].
    \eeq
We see that the scale anomaly of $\tr\log[-\grad^2 + \Si]$
exactly cancels that of $\int d^4x W_0(\Si)$. Therefore the
large-$N$ effective action
(\ref{e-eff-ac-large-N-finite-and-scale-invariant}) (for $m=0$)
is invariant under scale transformations
(\ref{e-scale-transformation}): $\delta_D \Gamma_0(B,\Si) = 0$.
However, $\G_0$ with a mass term
(\ref{e-eff-ac-large-N-finite-and-scale-invariant}) is not
scale-invariant. In fact,
    \beq
    \gd_D \G_0(B,\Si) = -2 \int d^4x~ m^2~ \Si(x) = -m \dd{}{m} \int d^4x~
    m^2~  \Si(x).
    \eeq
Define $\gb^m_0 =m$. Though $\G_0$ is not scale-invariant, it
satisfies a `renormalization group equation'
    \beq
    (\gd_D + \gb^m_0 \dd{}{m} ) \G_0(B,\Si) =0.
    \label{e-REG-large-N}
    \eeq
$\gb^m_0$ is the $N=\infty$ `beta function' of mass. In general, $\gb^m =
\gb^m_0 + \N \gb^m_1 + \Nsq \gb^m_2 + \cdots$. We could also have
$\gb^\la$ at higher orders in $1/N$, but we want
to ensure $\gb^\la =0$ for at least one value of $\la$ when $m=0$.
Let us define a new operator $\gd_0$, the large-$N$ `RG vector field':
    \beq
    \gd_0 = -x^\mu \dd{}{x^\mu} + b(x) \dd{}{b(x)} + 2 \si(x)
    \dd{}{\si(x)} + \gb^m_0 \dd{}{m}, ~~~~ {\rm where}~~~~ \gb^m_0 =
    m.
    \label{e-RGE-vfld-large-N}
    \eeq
Then $\gd_0 \Gamma_0(B,\Si) = 0$. The RG vector field may also receive corrections: $\gd = \gd_0 + {\gd_1 \over N} +
{\gd_2 \over N^2} + \cdots$.

\section{Small Oscillations around constant classical solutions}
\label{s-small-osc}

\subsection{Constant extrema of $N=\infty$ effective action}
\label{s-const-extrema-of-eff-action}

Field configurations that extremize the $N=\infty$ effective
action $\Gamma_0(B,\Si)$
(\ref{e-eff-ac-large-N-finite-and-scale-invariant}) dominate
the path integral over $b$ and $\si$ in the saddle point
approximation in $1/N$. Extrema of $\Gamma_0$ must satisfy the large-$N$
`classical' equations of motion (also called gap equations
elsewhere)
    \beq
    \fr{\delta \Gamma_0}{\delta B} &=& (- \grad^2 + \Si) B = 0 {\rm ~~~and} \cr
    \fr{\delta \Gamma_0}{\delta \Si} &=& \half (B^2 - m^2)
        + \bigg(\la - \fr{\hbar}{16\pi^2} \Pi(\Delta) \bigg) \Si +
        {\cal O}(\Si - \Si_o)^2  = 0.
    \label{e-class-eqn-motion}
    \eeq
Unlike, say, the gap equations of the non-linear sigma model in
the large-$N$ limit, these equations are finite and do not
require any renormalization, since the divergences in $\tr
\log[-\grad^2 + \Si]$ have been canceled by the choice of
$W_0(\Si)$. The complication here is that the equation of
motion for $\Si$ has been obtained in a series in powers of
$\Si - \Si_o$ for constant $\Si_o$. Let us begin by looking for
constant extrema ($B = B_o, \Si = \Si_o$) of $\Gamma_0$. The
classical equations of motion become
    \beq
    \Si_o B_o = 0   {\rm ~~and~~}  B_o^2 - m^2 + 2\la \Si_o  = 0 .
    \label{e-class-eqn-motion-constant-flds}
    \eeq
Assuming $\la$ and $m$ are non-zero, which is the generic
situation, there are two types of extrema: ({\bf S}) $B_o =0,
\Si_o = m^2/2\la$ where $O(N+1)$ symmetry is unbroken and ({\bf
B}) $B_o = \pm m$ and $\Si_o =0$ where $O(N+1)$ symmetry is
spontaneously broken to $O(N)$. The vev of the real scalar
field $b$ must be real. This justifies our choice for the sign
of the $m^2 \si$ term in $W_0(\si)$ in (\ref{e-W_0}).
In the broken phase, the vev $\Si_o = 0$. This is fine for constant backgrounds, since the effective potential is entire. But we cannot yet analyze the vicinity of the broken vacuum for non-constant backgrounds, since we found the effective action only in an expansion in powers of
$\Si - \Si_o$ which involves inverse powers of
$\Si_o$. Perhaps a cleverer method can be invented to study the
phase where $O(N+1)$ symmetry is broken. There may also be
phases where fields are not constant. Now,
we study oscillations around the vacuum where $O(N+1)$ symmetry
is unbroken.

\subsection{Mass of long wavelength small oscillations in unbroken phase}
\label{s-mass-of-lightest-particles}

To study oscillations around an extremum of the large-$N$ effective
potential, we expand the large-$N$ effective action
(\ref{e-eff-ac-large-N-finite-and-scale-invariant}) to quadratic
order around a constant background $\B = \B_o + \beta, \Si = \Si_o +
\vsi$:
    \beq
    \Gamma_0(\B_o + \beta,\Si_o + \vsi) &=& \half \int d^4x \bigg[ (\grad \beta)^2
        + \bigg\{2 \Si_o \B_o \beta + (\B_o^2 - m^2 + 2 \la \Si_o) \vsi \bigg\}
        \cr && +~ \bigg\{\Si_o \beta^2 + 2 \B_o \beta \vsi + \vsi \bigg(\la
        - \fr{\hbar}{16\pi^2} \Pi(\Delta)\bigg) \vsi \bigg\}
        +  \cdots \bigg].
    \eeq
We have omitted an additive constant in $\G_0$. Assuming that
$(\B_o,\Si_o)$ is a constant solution of the classical
equations of motion (\ref{e-class-eqn-motion-constant-flds}),
the linear terms drop out and we get
    \beq
    \Gamma_0(\B_o+\beta,\Si_o + \vsi) = \half \int d^4x
        \bigg[(\grad \beta)^2 - \fr{\hbar}{16\pi^2} \vsi \Pi(\Delta) \vsi
        + \Si_o \beta^2 + 2 \B_o \beta \vsi + \la \vsi^2 + \cdots \bigg].
    \label{e-gamma-0-exp-to-second-order}
    \eeq
Here $\Delta = - \grad^2/\Si_o$ and $\Pi(\Delta)$ is given in
(\ref{e-inv-sigma-propagator}). The lightest `particles' of the
theory are the longest wavelength oscillations around the
classical vacua. To find the mass of the longest wavelength
oscillations we may assume that the deviation $\vsi = \Si -
\Si_o$ is slowly varying in space. It suffices to keep the
leading term in $\Pi(\Delta) = -{\Delta \over 6} + {\Delta^2
\over 24} - {\Delta^3 \over 60} + \cdots $. Thus
    \beq
    \Gamma_0(\B_o+\beta,\Si_o + \vsi) = \half \int d^4x
        \bigg[(\grad \beta)^2 + \fr{\hbar}{96\pi^2 \Si_o} \vsi (-\grad^2) \vsi
        + \Si_o \beta^2 + 2 \B_o \vsi \beta + \la \vsi^2 + \cdots \bigg].
    \eeq
This should describe small amplitude oscillations since we also ignored higher powers of $\vsi$.
In symmetric phase ({\bf S}) where $O(N+1)$ is unbroken, the
classical minimum is at $\B_o = 0$, $\Si_o = m^2/2\la$.
$\Gamma_0$ expanded around this minimum is
    \beq
    \Gamma_0(\beta,\vsi) = \half \int d^4x \bigg[ (\grad \beta)^2
        + \fr{\hbar \la}{48 \pi^2 m^2} (\grad \vsi)^2  +
        \fr{m^2}{2\la} \beta^2 + \la \vsi^2 \bigg].
    \eeq
Recall that a field $\phi$ with Lagrangian $(\grad \phi)^2 + m^2
\phi^2$ has as its longest wavelength excitation, a particle of mass
$m$. We deduce that the lightest particle-like excitation of the $b$
field has a mass $M_b = \fr{m}{\sqrt{2\la}}$ and transforms in the fundamental representation of $O(N+1)$. Small
oscillations of $\si$ correspond to a particle of mass
$M_\si = \fr{4 \sqrt{3} \pi m}{\sqrt{\hbar}}$.
This particle is an $O(N+1)$ singlet. There could
also be heavier particles.



\section{Leading $1/N$ corrections due to fluctuations of $b$ and $\si$}
\label{s-leading-1-ov-N-corrns}

At $N=\infty$, quantum fluctuations of $b$ and $\si$ could be ignored,
but averaging over them at ${\cal O}(1/N)$ will lead to
divergences and violations of scale-invariance in the effective action $\Gamma$.
The technical difficulty in obtaining $\G(B,\Si)
= \G_0(B,\Si) + \N \G_1 (B,\Si)$ is that $\G_0(B,\Si)$
(\ref{e-eff-ac-large-N-finite-and-scale-invariant}), when expanded,
involves arbitrarily high powers and derivatives of $\Si$,
    \beq
        S_0(b,\si) = \Gamma_0(b,\si) = \half \int d^4x \bigg[
            (\grad b)^2 + \si b^2 - m^2 \si + \la \si^2 -\ov{16
            \pi^2} \bigg\{\si \Pi(\Delta) \si  + {\cal O}(\si - \si_o)^3 \bigg\}
        \bigg].
    \eeq
$\Pi(-\grad^2/\si_o)$ is defined in (\ref{e-inv-sigma-propagator}). The
action of our model at this order is
    \beq
        S(b,\si) = S_0(b,\si) + \ov{2N} \int d^4x W_1(\si)
            + {\cal O}(\ov{N^2}).
    \eeq
We must pick $W_1(\Si)$ so that $\Gamma_1(B,\Si)$ is finite and scale-free for $m = 0$.
However, this is most likely not possible since $W_1(\Si)$ is
independent of $B$ and can at best cancel divergences in
$\G_1(0,\Si)$. The remaining divergences must be
canceled via mass and coupling constant renormalization and
anomalous dimensions. These will modify the generator of
RG flow
    \beq
        \delta = \delta_0 + \N \delta_1 + \ov{N^2} \delta_2
            + \cdots; ~~~~
    \gd_1 = - \gb^m_1 \dd{}{m} - \g^B_1 B \dd{}{B} - \g^\Si_1 \Si
    \dd{}{\Si} - \gb^\la_1 \dd{}{\la}
    \eeq
such that $\Gamma$ still satisfies the
RGE $\delta \Gamma(B,\Si) = 0$ at
each order. In order to have a fixed
point, which would make it natural to have small masses, we
will impose $\beta^\la_1 =0$ for at least one value of $\la$ when $m=0$.
Since $\gd_0 \G_0 = 0$
(\ref{e-REG-large-N}), the RGE for the effective action at
order $1/N$ reads
    \beq
        \delta_0 \Gamma_1 + \delta_1 \Gamma_0 = 0.
    \label{e-RGE-at-order-1-over-N}
    \eeq
$\delta_0 = -x^\mu \dd{}{x^\mu} + b \dd{}{b} + 2 \si \dd{}{\si}
+ m \dd{}{m} $ and $\Gamma_0$
(\ref{e-eff-ac-large-N-arbit-bkgrnd}) are known, so we must
determine $\gd_1$ and $\G_1$.

\subsection{Calculation of effective action at ${\cal O}(1/N)$}
\label{s-eff-ac-at-1-ov-N}

Using (\ref{e-implicit-def-of-eff-action}) and doing the
gaussian integral at order $1/N$ we get the change in effective
action
    \beq
        \Gamma_1(B,\Si) = \half \bigg[ \tr \log \Gamma_0^{\prime
            \prime}(B,\Si) + \int d^4x~ W_1(\Si) \bigg].
    \label{e-gamma-1-basic-formula}
    \eeq
$\Gamma_0^{\prime \prime}(B,\Si)$ is the hessian of $\Gamma_0$
acting on the $2$-component column vector $( b ~~\si)$,
    \beq
        \Gamma_0^{\prime \prime}(B,\Si) = \left( \begin{array}{cc}
        -\grad^2 + \Si & B \\ B & - \fr{\Pi(-\grad^2/\Si_o)}{16\pi^2} + \la \\
        \end{array} \right)
    \label{e-hessian}
    \eeq
$\G_1$ is independent of $m$ even if $\G_0$ includes a mass
deformation.

\subsubsection{$W_1(\Si)$ for slowly varying quantum fluctuations \& general background $\Si(x)$}
\label{s-gamma-1-for-B-equal-0}

To find $W_1(\Si)$, we calculate $\G_1$
for $B=0$ and choose $W_1(\Si)$ to make $\G_1(0,\Si)$ finite
and scale-free for arbitrary $\Si(x)$.
(However, even for $B=0$, $\Si$ acquires an anomalous dimension, see sec. \ref{s-RGE-for-gamma-1}.)
For $B=0$, the hessian (\ref{e-hessian}) is diagonal, so
    \beq
    \Gamma_1(0,\Si) = \half\bigg[\tr \log[-\grad^2
            + \Si] + \tr \log\bigg[-\fr{\Pi(-\grad^2/\Si_o)}{16\pi^2}
        + \la \bigg] + \int W_1(\Si) d^4x \bigg].
    \eeq
Since $\Pi(-\grad^2/\Si_o)$ is complicated
(\ref{e-inv-sigma-propagator}), we make the further
approximation that quantum fluctuations are slowly
varying\footnote{Quantum fluctuations need not be slowly varying.
We hope to relax this assumption in future work.}. Then we may ignore higher derivatives in $\Pi(-\grad^2/\Si_o)$ and get
    \beq
    \Gamma_1(0,\Si) = \half\bigg[\tr \log[-\grad^2
            + \Si] + \tr \log[-\fr{\grad^2}{96\pi^2 \Si_o} + \la]
        + \int W_1(\Si) d^4x \bigg].
    \eeq
The $1^{\rm st}$ term is identical to what
appeared in $\Gamma_0$ (\ref{e-def-of-eff-ac-at-infinite-N})
and was calculated in an expansion around a constant background
$\Si_o$ in (\ref{e-tr-log-arbit-bkgrnd}). The $2^{\rm nd}$ term
can be calculated exactly in zeta function regularization.
Let $\zeta(s) = \tr
[-\grad^2 + 96 \pi^2 \la \Si_o]^{-s}$, then using (\ref{e-zeta-of-s-const-bkgrnd}) we get,
    \beq
    \tr \log[-\fr{\grad^2}{96\pi^2 \Si_o} + \la] &=& - \zeta(0)
        \log [96 \pi^2 \Si_o] - \zeta^{\prime}(0) \cr
    {\rm where ~~~} \zeta(0) = \fr{(96 \pi^2)^2 \la^2 \Si_o^2 \Om}{32 \pi^2}
        &{\rm and}&  \zeta^\prime(0) = \fr{(96 \pi^2)^2 \la^2 \Si_o^2 \Om}{32 \pi^2}
        \log[e^{-3/2} 96 \pi^2 \la \Si_o] \cr
    \implies    \tr \log[-\fr{\grad^2}{96\pi^2 \Si_o} + \la] &=&
        - \fr{(96\pi^2)^2 \la^2 \Si_o^2 \Om}{32 \pi^2}
        \log[e^{-3/2} (96\pi^2)^2 \la \Si_o^2].
    \eeq
$\la \Si_o$ cannot be a negative real number. So $\la =0$ is singular within our
approximations. Thus
    \beq
    \Gamma_1(0,\Si) = \half \bigg[\tr \log[-\grad^2 + \Si]
        - \fr{(96\pi^2)^2 \la^2 \Si_o^2 \Om}{32 \pi^2}
        \log[e^{-3/2} (96\pi^2)^2 \la \Si_o^2/M^4] + \int W_1(\Si) d^4x \bigg].
    \eeq
The $1^{\rm st}$ and $2^{\rm nd}$ terms violate scale-invariance and involve a
dimensional parameter $M$ as in section
\ref{s-scale-inv-of-effac-inf-N}. So
$W_1(\Si)$ is determined by the condition that it must cancel
these scale anomalies. Aside from the $2^{\rm nd}$ term, we had the
same condition for $W_0(\Si)$ in
(\ref{e-eff-ac-large-N-arbit-bkgrnd}). So\footnote{We could add a finite term
$\la_1 \Si^2$ to $W_1$ without violating scale invariance. We
did this for $W_0$, but do not for the other $W_n$, just as we
do not add an arbitrary finite counter term $\la_n \phi^4$ at
$n$-loop order of $\la \phi^4$ theory. In a sense, we allow the
most general $W_0$ consistent with the symmetries but the
remaining ones are the minimal choices that ensure cancelation
of divergences and preservation of scale invariance.} as in (\ref{e-W_0}),
    \beq
    W_1(\Si) &=& - \ov{16\pi^2}
        \bigg\{ \half \Si_o^2 \log[{\Si_o e^{-3/2} \over M^2}] + \Si_o
        \log[{\Si_o \over M^2 e}] (\Si - \Si_o) + \half \log[{\Si_o \over M^2}] (\Si - \Si_o)^2
        \cr &&
    - {(96\pi^2)^2 \la^2 \Si_o^2 \over 2} \log[{(96\pi^2)^2 \la \Si_o^2 \over e^{3/2} M^4}] \bigg\}.
    \label{e-W_1}
    \eeq
$W_0 + \N W_1$ is the potential for which $\Gamma_0 + \N \Gamma_1$ is finite and scale-free for
backgrounds $\Si(x)$ and $B=0$, after including slowly varying
quantum fluctuations at ${\cal O}(1/N)$. For this choice,
    \beq
    \G_1(0,\Si) = \half \int d^4x \bigg[-\ov{16 \pi^2} (\Si -\Si_o) \Pi(-\grad^2/\Si_o) (\Si-\Si_o)
        + {\cal O}(\Si-\Si_o)^3 \bigg].
    \eeq

\subsubsection{RGE for slowly varying fluctuations and constant backgrounds}
\label{s-RGE-for-gamma-1}

In section \ref{s-gamma-1-for-B-equal-0} we found $W_1(\Si(x))$
that ensures $\G_1$ is finite and scale-free for $B=0$. But
there could be further divergences which lead to running
couplings and anomalous dimensions. We determine $\gb^m$,
$\g^\si$ and $\g^b$, while enforcing $\gb^\la=0$ for constant
backgrounds $B_o$ \& $\Si_o$. We do this by deriving an RGE for
the effective action. We assume quantum fluctuations of $\si$
are slowly varying on the scale of the constant background
$\Si_o$, so that
    \beq
        \Gamma_0^{\prime \prime}(\B_o,\Si_o) = \left(\begin{array}{cc}
            -\grad^2 + \Si_o & \B_o \\
            \B_o & - \fr{\grad^2}{96 \pi^2 \Si_o} + \la \\
            \end{array} \right) = \left(\begin{array}{cc}
            A & B_0 {\bf 1} \\
            B_0 {\bf 1} & D \\
            \end{array} \right).
    \eeq
Here $A = -\grad^2 + \Si_o$, $D = - \grad^2/\tilde \Si_o + \la$ and
$\tilde \Si_o = 96 \pi^2 \Si_o$. Since $B_0 {\bf 1}$ commutes with
$D$, $\det[\Gamma_0^{\prime \prime}] = \det[AD - B_o^2 {\bf 1}]$,
i.e., the characteristic polynomial of $AD$. To calculate it, let
    \beq
    \zeta(s) = \tr[AD - B_o^2]^{-s} = \fr{\Om}{(2\pi)^4} \int d\Om_4 \int_0^\infty
        \fr{p^3 dp}{ \bigg(p^4/\tilde \Si_o + (\Si_o/\tilde \Si_o + \la) p^2
        + (\la \Si_o - B_o^2)\bigg)^{s} }.
    \eeq
The integral converges for $\Re{s} > 1$ and defines
meromorphic $\zeta(s)$. Changing variables to $q = p^2$ and defining $2c =
\Si_o + \la \tilde \Si_o$, $d = (\la \Si_o - B_o^2) \tilde
\Si_o$, $\zeta(s)$ is expressed as
    \beq
    \zeta(s) = \fr{\Om \tilde \Si_o^{s}}{16\pi^2} \int_0^\infty \fr{q~ dq}
        {(q^2 + 2c q+ d)^s}
    = \fr{\Om \tilde \Si_o^{s}}{32\pi^2} \bigg[\fr{d^{1-s}}{s-1} -
        \fr{c^{2-2s}~ {_2 F_1}(s-\half,s,s+\half;1-d/c^2)}{(s-\half)}
        \bigg].
    \eeq
$\zeta(s)$ is analytic at $s = 0$ and we define $\tr\log[AD-B_o^2] = -
\zeta^{\prime}(0)$. Then from (\ref{e-gamma-1-basic-formula})
    \beq
    \Gamma_1(B_o,\Si_o) = -\half \zeta^\prime(0) + \half \Om
    W_1(\Si_o).
    \eeq
$\zeta^{\prime}(0)$ is complicated, but to understand the RGE
$\gd_0 \G_1 + \gd_1 \G_0 =0$  (\ref{e-RGE-at-order-1-over-N}), we
only need
    \beq
    \zeta(0) = \fr{\Om}{32 \pi^2} (2c^2 - d)
        = \fr{\Om}{64 \pi^2} \bigg[\Si_o^2 + \la^2 \tilde \Si_o^2 + 2 B_o^2
            \tilde \Si_o\bigg]
     = \fr{\Om}{64 \pi^2} \bigg[\{1 + (96\pi^2)^2 \la^2\} \Si_o^2
         + 192 \pi^2 B_o^2 \Si_o \bigg].
    \eeq
Recall from Sec.~\ref{s-cancel-scale-anomaly} that $\delta_0
\zeta^\prime(0) = -2 \zeta(0)$. Therefore,
    \beq
        \delta_0 \Gamma_1 = \delta_0~\bigg(-\half \zeta^\prime(0) + \half \Om
            W_1(\Si_o)\bigg)
        = \zeta(0) + \half \delta_0 ~\Om W_1(\Si_o).
    \eeq
Thus, the RGE (\ref{e-RGE-at-order-1-over-N}) for the effective action
becomes
    \beq
    \zeta(0) + \half \delta_0~ \Om W_1(\Si_o) = - \delta_1~
        \Gamma_0.
    \eeq
From (\ref{e-gamma0-zeta-fn-const-bkgrnd}) we know that
$\G_0(B_o,\Si_o) = \fr{\Om}{2} [\Si_o B_o^2 - m^2 \Si_o + \la
\Si_o^2 ]$. Let us parameterize the leading $1/N$ correction to
the renormalization group vector field as
    \beq
    \delta_1 = - \beta^m \dd{}{m} -
    \gamma^b B_o \dd{}{B_o} - \gamma^\si \Si_o \dd{}{\Si_o}.
    \eeq
$\beta^m(\la,m)$ is the beta function\footnote{$\beta$ and
$\gamma$ may receive corrections at ${\cal O}(1/N^2)$ as part
of $\delta_2$, so $\beta^m$ is short for $\beta^m_1$ etc.} of
$m$ while $\gamma^{b,\si}(\la,m)$ are anomalous dimensions. We
are explicitly imposing the condition $\beta^\la =0$, i.e. that
$\la$ remains unrenormalized and we have a line of fixed
points. We only need one fixed point for naturalness, but as
the sequel shows, we find a line of them within our approximations.
$W_1(\Si_o)$, $\beta$ and $\gamma$ must satisfy
    \beq
    \zeta(0) + \half \delta_0 \Om W_1(\Si_o) &=& \bigg[ \beta_m \dd{}{m}
        + \gamma^b B_o \dd{}{B_o}
        + \gamma^\si \Si_o \dd{}{\Si_o} \bigg] \fr{\Om}{2} \bigg(
        \Si_o B_o^2 - m^2 \Si_o + \la \Si_o^2 \bigg) \cr
    \Rightarrow ~~ \half \delta_0 \Om W_1(\Si_o) &+&
        \fr{\Om}{64 \pi^2} \bigg[\{1 + (96\pi^2)^2 \la^2\} \Si_o^2
         + 192 \pi^2 B_o^2 \Si_o \bigg]  \cr
    &=& \fr{\Om}{2} \bigg[(-2m \beta^m - m^2 \gamma^\si) \Si_o
        + 2\la\gamma^\si \Si_o^2 + (2 \gamma^b + \g^\si) \Si_o B_o^2 \bigg].
    \label{e-RGE-explicitly}
    \eeq
We already determined $W_1(\Si)$ in sec.
\ref{s-gamma-1-for-B-equal-0}, from it we calculate
    \beq
    \half \gd_0 \Om W_1(\Si_0) = -{\Om \Si_o^2 \over 32 \pi^2}
        + {(96 \pi^2)^2 \la^2 \Om \Si_o^2 \over 16 \pi^2}.
    \eeq
Putting this in (\ref{e-RGE-explicitly}) we get
    \beq
    \bigg(720 \pi^2 \la^2 - \ov{64 \pi^2} \bigg) \Si_o^2 + 3 B_o^2 \Si_o
        = \ov{2} \bigg[(-2m \beta^m - m^2 \gamma^\si) \Si_o
        + 2\la\gamma^\si \Si_o^2 + (2 \gamma^b + \g^\si) \Si_o B_o^2
        \bigg].
    \eeq
This equation must hold for {\em all} constant backgrounds
$B_o$ and $\Si_o$. If our model is not renormalizable or does not have a line of fixed points (for
slowly varying quantum fluctuations at ${\cal O}(1/N)$), there
would be no choice of $\beta^m, \g^{b,\si}$ for which it is
identically satisfied. Comparing coefficients of monomials in
the fields, we get the relations
    \beq
        2m \beta^m + m^2 \gamma^\si = 0, ~~~
        \la \gamma^\si = 720 \pi^2 \la^2 - \ov{64\pi^2},
        ~~~~ {\rm and ~~~}
        \gamma^b + \half \g^\si = 3,
    \eeq
whose unique solution is\footnote{As noted earlier, $\la=0$ is
a singular limit within our approximations, this is reflected
in $\gb^m, \g^\si, \g^b$.} (notice that $\gb^m \propto m$ so that
$m=0$ is preserved by RG flow)
    \beq
        \beta^m = m \bigg(\ov{128 \pi^2 \la} - 360 \pi^2 \la \bigg); ~~~
        \g^\si = 720 \pi^2 \la - \ov{64 \pi^2 \la}, ~~~
        \g^b = 3 + {1 \over 128 \pi^2 \la} - 360 \pi^2 \la.
    \eeq
The resulting RG vector field is
    \beq
    \gd &=&  \bigg\{1 + \N \bigg( 360\pi^2 \la - \ov{128 \pi^2 \la} \bigg) \bigg\} m \dd{}{m}
        ~+~ \bigg\{1 + {1 \over N} \bigg( 360 \pi^2 \la -3 - {1 \over 128 \pi^2 \la} \bigg) \bigg\} B_o \dd{}{B_o} \cr
        &&
        +~ \bigg\{ 2 + \N\bigg( {1 \over 64 \pi^2 \la} - 720 \pi^2 \la \bigg) \bigg\} \Si_o \dd{}{\Si_o}
        ~+~ {\cal O}(1/N^2).
    \eeq
For constant backgrounds and slowly varying quantum
fluctuations of $\si$, we have a consistent solution of the RGE
for which $\la$ remains unrenormalized ($\beta^\la =0$). Under
these assumptions, our model is renormalizable at order $1/N$.
It remains to see whether renormalizability and
$\beta^\la=0$ can be maintained (for at least one $\la$), for
non-constant backgrounds and rapidly varying quantum
fluctuations, perhaps via an expansion in inverse powers of
$\Si_o$.


\section*{Acknowledgements}

We thank S. G. Rajeev and G. 't Hooft for discussions on fixed
points and naturalness. We also acknowledge conversations with E. Laenen, E. Sezgin, O. T. Turgut, R.
Boels, R. K. Kaul and B. Sathiapalan. We thank the EU for support in the form of a Marie Curie
fellowship. This work was partially supported by EU-RTN network
MRTN-CT-2004-005104 {\em Constituents, Fundamental Forces and
Symmetries of the Universe.}

\appendix

\section{Examples of naturalness}
\label{a-eg-naturalness}

By a naturalness explanation for an `unreasonably small'
quantity we mean that the model acquires an additional symmetry
when that quantity vanishes \cite{tHooft-naturalness}. In the
absence of such a symmetry, if the quantity is dimensionless,
the `reasonable' or natural value for it is of order $1$, and
if it has dimensions, its natural value is of the order of the
Planck scale, though there may be another scale depending on the context. The symmetry used in a
naturalness explanation may be either continuous or discrete.
Sometimes, it is more convenient to refer to the conservation
law that follows from the symmetry. The actual small value of
the quantity (if non-zero) is usually related to explicit
breaking of the symmetry and can often be treated
perturbatively. We give a few examples of naturalness
explanations from different branches of physics. It appears
that this concept explains several
small parameters both in experimentally tested theories and
mathematical models. Indeed, except for the Higgs mass, there
is a naturalness explanation for every small parameter in the standard model.
This gives us some confidence to turn
things around: if there is an unreasonably small parameter in
nature or in a model, then there must be some symmetry, which
if exact, would make that parameter vanish. In this way,
naturalness can be a useful guide to model building.

\vs{-3}

{\fl \bf (1)} Small electron mass compared to Planck mass: If
$m_e =0$ QED would gain chiral symmetry. The
same applies to muon and tau masses. There is a different
chiral symmetry for each. This puts no
constraint on the ratios of lepton masses.

\vs{-3}

{\fl \bf (2)} Small current quark masses: For $N_{flavors} > 1$
if current quark masses are set to zero,
QCD gains a partial unbroken chiral symmetry $SU(N_f)_{V}$ (isospin for $N_f=2$).

\vs{-3}

{\fl \bf (3)} Small coupling constants can be explained
by the separate conservation laws
for particles, gained by setting coupling
constants to zero.

\vs{-3}

{\fl \bf (4)} Some near-degeneracies of energy levels in atomic
spectroscopy can be explained as due to the presence of a
symmetry. In hydrogen-like atoms, the difference in energy
between levels with the same principal and angular momentum
quantum numbers $n,l$ but different magnetic quantum number $m$
vanishes if we have spherical symmetry. The small energy
difference can be due to a magnetic field whose direction
explicitly breaks spherical symmetry.

\vs{-3}

{\fl \bf (5)} Naturalness in classical mechanics: The fact that
planetary orbits are nearly closed and nearly lie on a plane,
are related to the conservation of angular momentum and the
Laplace-Runge-Lenz vector in the Kepler problem. Quantum
mechanically, in hydrogen-like atoms, the `accidental
degeneracy' of energy levels with the same value of $l$ is due
to a hidden $SO(4)$ symmetry whose conserved quantities are
angular momentum and Laplace-Runge-Lenz vectors.

\vs{-3}

{\fl \bf (6)} Some near-degeneracies in atomic energy levels
can be explained as due to parity symmetry. The small
splittings are due to parity violation in the weak
interactions.

\vs{-3}

{\fl \bf (7)} Experimentally, the mass of a photon is less than
$10^{-16}$ eV outside a superconductor. This is explained by
the exact $U(1)$ gauge symmetry if the photon is massless.

\vs{-3}

{\fl \bf (8)} The near degeneracy in the proton-neutron masses
(and pion masses) may be explained as a consequence of isospin
symmetry. If $u$ and $d$ quarks were degenerate in mass,
isospin would be an exact symmetry of the strong interactions
and the neutron and proton would be degenerate in mass (as
would the three pions). Isospin is explicitly broken by the
quark mass difference as well as electromagnetic interactions,
which explain the small $n-p$ and $\pi^+-\pi^0$ mass
differences.

\vs{-3}

{\fl \bf (9)} Pions are naturally light compared to say,
$\rho$ mesons due to chiral symmetry. They are
the pseudo-goldstone bosons of spontaneously broken chiral
symmetry. If the quarks were massless, chiral symmetry would be
exact at the level of the lagrangian, and be spontaneously
broken to ${SU(N_f)}_V$, and the pions would be massless goldstone
bosons. But in fact, quarks are not massless, this explicit
breaking of chiral symmetry gives the pions a small mass.

\vs{-3}

{\fl \bf (10)} Small neutrino masses: Chiral symmetry for each
flavor is exact if neutrinos are massless.

\vs{-3}

{\fl \bf (11)} Parity is an exact symmetry of QCD in the
absence of the topological $\theta$ term, which is parity odd.
Thus, a small QCD theta angle is natural within the theory of
strong interactions.

\vs{-3}

{\fl \bf (12)} The scalar field mass in a supersymmetric
extension of the standard model can be naturally small since if
it were zero, the theory would have unbroken global
supersymmetry (when the super partner fermion is also massless,
which would be natural due to chiral symmetry).

\vs{-3}

{\fl \bf (13)} The effective mass (inverse of
correlation length) is very small in the
neighborhood of a $2^{\rm nd}$ order phase transition.
A naturalness explanation is that at such a transition, when
the effective mass vanishes, the system gains a new symmetry,
scale-invariance. However, one might argue that a naturalness
explanation is not needed here, since the temperature must be
fine-tuned in order to have such a phase transition.

\vs{-3}

{\fl \bf (14)} Some linear combinations of correlations in
large-$N$ multi-matrix models vanish because of the presence of
hidden non-anomalous symmetries \cite{ward-id-sde}.

\vs{-3}

{\fl \bf (15)} It has been suggested \cite{tHooft-nobbenhuis}
that a discrete symmetry that relates real-valued space-time
coordinates to pure-imaginary ones could ensure a naturally
small cosmological constant.

\vs{-3}

{\fl \bf (16)} The amplitudes for gluon scattering either with
the same helicity or with only one gluon with a different
helicity from the others, vanishes in classical Yang-Mills
theory. A naturalness explanation for this has been suggested,
using an effective tree-level supersymmetry
\cite{Dixon:1996wi}. This is an example of naturalness in
classical field theory.

\vs{-3}

{\fl \bf (17)} GIM mechanism and suppression of flavor-changing
neutral currents (FCNC) \cite{richter}: The small quantity here
is the $\D s =1$ strangeness changing neutral current compared
to the $\D s = 0$ neutral current. The GIM mechanism eliminates
$\D s=1$ FCNC at tree level by introducing a new quark doublet
participating in the weak interactions, consisting of charm and
the Cabibbo rotated strange quark. This is not a naturalness
explanation, but could be turned into one by specifying a
symmetry which ensures absence of FCNC.

\section{The quantum effective action} \label{a-eff-action}

To be self-contained, we collect a few facts about the effective action (see eg. \cite{nair-QFT}).
The generating series for (possibly disconnected)
correlations of a scalar field $\phi$ is
    \beq
        Z(J) = \int [D\phi] e^{-\ov{\hbar}[S(\phi) + J \cdot \phi] } {\rm ~~~~where~~~} J \cdot \phi = \int d^4x~ J(x) \phi(x)
    \eeq
The generating series of connected correlations is $W(J) = - \hbar \log Z(J)$.
The effective action $\Gamma(\Phi)$ is the generating series of
proper vertices ($1$-particle irreducible diagrams). It is the
Legendre transform of $W(J)$: $\Gamma(\Phi) = {\rm ext}_{J}[W(J) - J \cdot \Phi]$.
The solution of the {\em classical} theory (tree diagrams) defined by
$\Gamma$ is equivalent to a solution of
the {\em quantum} theory of the original action $S$. An implicit integral
representation for $\Gamma(\Phi)$ is
    \beq
        e^{-\Gamma(\Phi)/\hbar} = \int [D \phi] e^{-\ov{\hbar}
            [S(\Phi + \phi) - \phi \cdot {\delta \Gamma(\Phi)
            \over \delta \Phi} ]}.
        \label{e-implicit-formula-for-eff-ac}
    \eeq
Here $\Phi$ is the background field and $\phi$ is the
fluctuating field. $\Phi$ need not solve the
classical equations of motion, i.e. $S^{\prime}(\Phi)$ need not
be zero. To obtain this integral representation we start from
the relation $W(J) = {\rm ext}_{\Phi} [\Gamma(\Phi) + J \cdot \Phi].$
The extremum occurs at $J = - {\delta \Gamma(\Phi) \over \delta
\Phi}$. With $J$ and $\Phi$ related this way, the effective
action is
    \beq
        \Gamma(\Phi) = W(J) - J \cdot \Phi = -\hbar \log Z(J) - J
            \cdot \Phi.
    \eeq
In other words, $e^{-\ov{\hbar} \Gamma(\Phi)} = Z(J) e^{\ov{\hbar}J \cdot
\Phi}$. Inserting the path integral for $Z(J)$,
    \beq
        e^{-\ov{\hbar} \Gamma(\Phi)} = \int [D\chi] e^{-\ov{\hbar} [S(\chi) + J
            \cdot \chi - J \cdot \Phi]}.
    \eeq
Changing variables of integration to $\phi = \chi - \Phi$ and using $J = J(\Phi) = - {\delta \Gamma(\Phi) \over \delta \Phi}$, we
get the implicit formula (\ref{e-implicit-formula-for-eff-ac}) for the
effective action.
By expanding $S(\Phi + \phi)$ in powers of $\phi$ and doing the
gaussian integrals, we get an asymptotic series for $\Gamma(\Phi)$ in powers of $\hbar$
(the same formula holds with $\hbar$ replaced by $1/N$)
    \beq
        \Gamma(\Phi) = S(\Phi) + {\hbar \over 2} \log \det{S^{\prime
        \prime}(\Phi)} + {\cal O}(\hbar^2).
    \eeq

\subsection{Large-$N$ effective potential via dimensional regularization}
\label{a-dim-reg}

We wish to calculate $\tr \log [-\grad^2 + \Si_o] = \Om \int
[d^4p] \log[p^2 + \Si_o]$ appearing in the large-$N$ effective
action (\ref{e-def-of-eff-ac-at-infinite-N}) via dimensional
regularization. Analytically continuing to $n$ Euclidean
dimensions and differentiating, we get a convergent integral if
$n < 2$
    \beq
    T_n = \int [d^n p] \log[p^2 + \Si_o] ~~\implies~~
    \dd{T_n}{\Si_o} = \int {[d^n p] \over (p^2 + \Si_o)} =
    (4\pi)^{-n/2}~ {\G(1-n/2) \over \Si_o^{1-n/2}}.
    \eeq
Here $[d^np] = {d^np \over (2\pi)^n}$. Expanding in a Laurent
series around $n=4$ dimensions,
    \beq
    \dd{T_n}{\Si_o} = {\Si_o \over 8 \pi^2 (n-4)} + {\Si (\g -1 + \log[\Si_o/4\pi]) \over 16
    \pi^2} + {\cal O}(n-4).
    \eeq
Now $\hbar \tr \log[-\grad^2 + \Si_o] = \hbar \Om T_n$. So
integrating with respect to $\Si_o$,
    \beq
    \hbar \tr \log[-\grad^2 + \Si_o] = {\hbar \Om \Si_o^2 \over 16 \pi^2
    (n-4)} + {\hbar \Om \Si_o^2 \over 32 \pi^2}(\g - {3 \over 2} - \log
    4\pi) + {\hbar \Om \Si_o^2 \log \Si_o \over 32 \pi^2} + c
    \hbar \Om + {\cal O}(n-4).
    \eeq
Integration constant $c$ is independent of $\Si_o$ and plays no
role since it only contributes an additive constant to the
effective potential. $\g=.577$ is Euler's constant. We have a
pole part, finite part and terms that vanish as $n \to 4$.
Notice that the finite part that transforms inhomogeneously
under scale transformations ${\hbar \Om \Si_o^2 \log \Si_o
\over 32 \pi^2}$ is the same as in cutoff or zeta-function
regularization. The choice of $W_0$ that makes the effective
action finite and scale-free in the limit $n \to 4$ is
    \beq
    W_0(\Si_o,n) = -{\hbar \Si_o^2 \over 16 \pi^2 (n-4)} - {\hbar \Si_o^2 \log \Si_o \over 32
    \pi^2} + \la \Si_o^2.
    \eeq
The finite and scale-free effective action for constant
backgrounds is thus
    \beq
    \G_0(B_o,\Si_o) = {\Om \over 2} \bigg[\bigg(\la
        + {\hbar (\g - 3/2 - \log 4\pi) \over 32 \pi^2} \bigg)
        + \Si_o^2 B_o \bigg].
    \eeq
We get a line of fixed points parameterized by $\la$, whose
definition is scheme dependent
    \beq
    \la_{\rm zeta-fn} = \la_{\rm dim-reg} + {\hbar (\g - 3/2 - \log 4\pi) \over 32
    \pi^2}.
    \eeq

\section{Expansion of $\tr \log[-\grad^2 + \si]$ in powers and derivatives
of $\si$}

\label{a-exp-for-tr-log}

\subsection{Zeta function in terms of the heat kernel}
\label{a-zeta-fn-heat-kernel}

Let $A = -\grad^2 + \si(x)$ and $\zeta_A(s) = \tr A^{-s}$.
Then $\tr \log A = \zeta'(0)$. We get an integral representation
for $\zeta_A(s)$ by making a change of variable $t \mapsto At$ in
the formula for $\Gamma(s)$:
    \beq
        A^{-s} = \ov{\Gamma(s)} \int_0^\infty dt ~e^{-tA}~ t^{s-1}.
    \eeq
Now define the evolution operator $\hat h_t = e^{-tA}$ which
satisfies a generalized heat equation
    \beq
        {d \hat h_t \over dt} = -A \hat h_t = ( \grad^2 - \si) ~\hat h_t, ~~~~ \lim_{t \to 0^+} \hat h_t = 1
    \eeq
It is convenient to work with the heat kernel $\hat h_t ~\psi(x) = \int ~dy~ h_t(x,y) ~\psi(y)$
which satisfies
    \beq
        \dd{h_t(x,y)}{t} = \bigg[\grad^2 - \si(x) \bigg] ~h_t(x,y) {~~~\rm and ~~~}
        \lim_{t \to 0^+} h_t(x,y) = \delta(x-y).
    \label{e-gen-heat-eqn}
    \eeq
Then $\zeta_A(s)$ is the Mellin transform of the trace of
the heat kernel:
    \beq
        \zeta_A(s) = \tr A^{-s} = \ov{\Gamma(s)} \int_0^\infty dt ~t^{s-1} ~\tr e^{-tA}
        = \ov{\Gamma(s)} \int_0^\infty dt ~t^{s-1} ~ \int d^4x
        ~h_t(x,x).
    \label{e-def-of-zeta-as-mellin-transf}
    \eeq
To find $h_t(x,x)$ we need to solve (\ref{e-gen-heat-eqn}). We find
$h_t(x,y)$ and take $x \to y$ in the end.
For $\si = 0$, $h_t(x,y)$ satisfies the diffusion equation and the
solution is
    \beq
        h^o_t(x,y) = \ov{(4\pi t)^2} e^{-{(x-y)^2 \over 4t}}.
    \eeq
For constant complex $\si = \si_o$
(\ref{e-gen-heat-eqn}) is a PDE with constant coefficients whose solution is
    \beq
        h^o_t(x,y) = \ov{(4 \pi t)^2} e^{-t \si_o} e^{-{(x-y)^2 \over
        4t}}.
    \label{e-heat-ker-const-sigma}
    \eeq


\subsection{Short time expansion for heat kernel}
\label{s-short-time-exp-heat-kernel}

For non-constant $\si$ we get an expansion for the heat kernel in
derivatives and powers of $\si$ for small $t$. We assume that the
non-analytic part of the heat kernel is already captured by the
exact solution (\ref{e-heat-ker-const-sigma}). We let $\si = \si_o +
\vsi(x)$ and make the ansatz
    \beq
        h_t(x,y) = h_t^o(x,y) ~\sum_{n=0}^\infty a_n(x,y) t^n  =
        {e^{-\si_o t} e^{-(x-y)^2/4t} \over (4\pi t)^2} ~\sum_{n=0}^\infty a_n(x,y)
        t^n.
    \label{e-ansatz-for-heat-kernel}
    \eeq
The average value of $\vsi$ need not vanish. However, we assume that
$\grad \vsi(x) \to 0$ as $|x| \to \infty$ so that the average value
of derivatives of $\vsi$ and its powers vanish
    \beq
        \int dx ~(\grad^2)^p \vsi^q(x) = 0, ~~~ p,q = 1,2,3 \ldots
    \eeq
For (\ref{e-ansatz-for-heat-kernel}) to satisfy the initial
condition (\ref{e-gen-heat-eqn}), we need $a_0 = 1$. If $\si$
is a constant, $a_i = \delta_{0,i}$. In
(\ref{e-ansatz-for-heat-kernel}), we could absorb $e^{-\si_o
t}$ into the infinite series since it is analytic in $t$, but
that would amount to throwing away an exact result, so let us
not do it. Moreover, $e^{-\si_o t}$ makes the Mellin
transform (\ref{e-def-of-zeta-as-mellin-transf}) of the heat
kernel convergent for $\Re \si_o > 0$, which is necessary to
recover $\zeta_A(s)$. The coefficients $a_n(x,y)$ are to be obtained by putting
(\ref{e-ansatz-for-heat-kernel}) into the heat equation
(\ref{e-gen-heat-eqn}) and comparing coefficients of common powers
of $t$. We need the expressions
    \beq
        \dd{h_t}{t} &=& \dd{h_t^o}{t} ~\sum_0^\infty ~a_n ~t^n
        ~+~ h_t^o ~\sum_1^\infty n ~a_n ~t^{n-1} ~~~~~ {\rm and}  \cr
        \grad^2 h_t &=& \grad^2~ h_t^o ~\sum_0^\infty ~a_n ~t^n
        +2 \grad_i h^o_t ~\sum_0^\infty ~t^n \grad_i ~a_n
        + h_t^o \sum_0^\infty ~ t^n \grad^2 a_n.
    \eeq
We put these into the generalized heat equation (\ref{e-gen-heat-eqn}) and get
    \beq
        \dd{h_t^o}{t} ~\sum_0^\infty ~a_n ~t^n
        ~+~ h_t^o ~\sum_1^\infty n ~a_n ~t^{n-1} &=&
        \grad^2~ h_t^o ~\sum_0^\infty ~a_n ~t^n
        +2 \grad_i h^o_t ~\sum_0^\infty ~t^n \grad_i ~a_n
        \cr && + h_t^o \sum_0^\infty ~ t^n \grad^2 a_n
        - (\si_o + \vsi) h_t^o \sum_0^\infty a_n t^n.
    \eeq
Using the fact that $\pdr_t h^o = (\grad^2 - \si_o) h^o$ this
simplifies to
    \beq
        \sum_0^\infty (n+1) ~a_{n+1} ~t^n =
        2 {\grad_i h^o_t \over h^o_t} ~\sum_0^\infty ~t^n \grad_i ~a_n
        + \sum_0^\infty ~ t^n \grad^2 a_n
        - \vsi \sum_0^\infty a_n t^n.
    \eeq
Now
    \beq
        {\grad_i h^o \over h^o} = \grad_i \log h^o = \grad_i
        \bigg[-(x-y)^2/4t \bigg]= -{(x-y)_i \over 2t}.
    \eeq
So the generalized heat equation becomes
    \beq
        \sum_0^\infty (n+1) ~a_{n+1} ~t^n =
        -(x-y)_i ~\sum_0^\infty ~t^n \grad_i ~a_{n+1}
        + \sum_0^\infty ~ t^n \grad^2 a_n
        - \vsi \sum_0^\infty a_n t^n.
    \eeq
Comparing coefficients of $t^n$ determines
$a_{n+1}$ given $a_n$ with the initial condition $a_0 = 1$
    \beq
        \bigg\{(x-y)_i + n + 1 \bigg\} a_{n+1}(x,y) = (\grad^2 - \vsi)
        a_n(x,y).
    \eeq
Now only $a_n(x,x)$ appear in $\zeta(s)$, so we specialize to
    \beq
        a_{n+1}(x,x) = \ov{(n + 1)}(\grad^2 - \vsi) a_n(x,x).
    \eeq
The first few $a_n(x,x)$ are
    \beq
        a_1 &=& -\vsi(x); ~~~~~
        a_2 = \half (\grad^2 - \vsi) a_1 = \half (\vsi^2 - \grad^2
            \vsi) \cr
        a_3 &=& \ov{3}(\grad^2 - \vsi) a_2 = \ov{3!} (\grad^2 - \vsi)
        (\vsi^2 - \grad^2 \vsi) = \ov{3!} (\vsi \grad^2 \vsi - \vsi^3
            + \grad^2 \vsi^2 - (\grad^2)^2 \vsi) \cr
        a_4 &=& \ov{4} (\grad^2 - \vsi) a_3 = \ov{4!} (\grad^2 -
            \vsi) (\vsi \grad^2 \vsi - \vsi^3 + \grad^2 \vsi^2 - (\grad^2)^2
            \vsi) \cr
        &=& \ov{4!} (\grad^2(\vsi \grad^2 \vsi) - \grad^2 \vsi^3 + \grad^4 \vsi^2
            - \grad^6 \vsi - \vsi^2 \grad^2 \vsi
            + \vsi^4 - \vsi \grad^2 \vsi^2 + \vsi \grad^4 \vsi).
    \eeq
To summarize, the heat kernel expansion is $h_t(x,x) = {e^{-\si_o t} \over (4\pi t)^2}
\sum_0^\infty a_n t^n$. If we drop cubic and higher powers of $\vsi$, then the $a_n$ are
    \beq
        a_0 ~=~ 1; ~~~
        a_1 ~=~ -\vsi; &&
        a_2 ~=~ \half (\vsi^2 - \grad^2 \vsi); ~~~
        a_3 ~=~ \ov{3!} (\vsi \grad^2 \vsi + \grad^2 \vsi^2 - (\grad^2)^2 \vsi) + {\cal O}(\vsi^3); \cr
        a_4 &=& \ov{4!} (\grad^2(\vsi \grad^2 \vsi) + \grad^4 \vsi^2 - \grad^6 \vsi + \vsi \grad^4
        \vsi) + {\cal O}(\vsi^3) {\rm ~~~~~ etc.}
    \eeq
For $\zeta_A(s)$ we need $\bra a_n \ket = \int d^4 x a_n(x,x) / \int
d^4 x$. Assuming $\vsi \to$ constant as $|x| \to \infty$ and $\grad
\vsi \to 0$ at $\infty$ we get (up to terms involving cubic and
higher powers of $\vsi$),
    \beq
        \bra a_0 \ket = 1, ~~~
        \bra a_1 \ket = - \bra \vsi \ket, ~~~
        \bra a_2 \ket = \ov{2!} \bra \vsi^2 \ket, ~~~
        \bra a_3 \ket = \ov{3!} \bra \vsi \grad^2 \vsi \ket + {\cal O}(\vsi^3) {\rm ~~~~etc.}
    \label{e-list-of-avg-val-of-a_n}
    \eeq
More generally, $\bra a_n \ket = \ov{n!} \bra \vsi (\grad^2)^{n-2}
\vsi \ket + {\cal O}(\vsi^3)$ for $n = 3,4,5,\ldots$.

\subsection{Derivative expansion for $\tr \log[-\grad^2 + \si]$}
\label{a-derivative-exp-of-tr-log}

We use
(\ref{e-def-of-zeta-as-mellin-transf}) and the expansion
(\ref{e-ansatz-for-heat-kernel}) to get an expansion for
$\zeta_A(s)$ in derivatives of $\si = \si_o + \vsi$
    \beq
        \zeta(s) &=& \ov{\Gamma(s)} \int_0^\infty dt
            \int d^4x ~t^{s-1} ~h_t(x,x) =
        \ov{\Gamma(s)} \int d^4x \int_0^\infty dt~ t^{s-1} {e^{-\si_o t}
            \over (4\pi t)^2} \sum_0^\infty a_n(x,x) t^n \cr
        {\zeta(s) \over \Om} &=& \ov{16\pi^2 \Gamma(s)} \sum_0^\infty \bra a_n \ket
            \int_0^\infty dt~ t^{s+n-3} e^{-\si_o t}
    \eeq
where $\Om = \int d^4x$. The integral over $t$ is a
Gamma function,
    \beq
        {\zeta(s) \over \Om} &=& \ov{16\pi^2 \Gamma(s)} \sum_n \bra a_n \ket
            \si_o^{2-n-s} \Gamma(s+n-2)
        = {\si_o^{2-s} \over 16 \pi^2} \sum_n {\bra a_n \ket \over \si_o^{n}}
            {\Gamma(s+n-2) \over \Gamma(s)} \cr
        &=& {\si_o^{2-s} \over 16\pi^2} \bigg[{\bra a_0 \ket \over
            (s-1)(s-2)} + {\bra a_1 \ket \over (s-1) \si_o} + {\bra a_2 \ket \over \si_o^2} +
            {s \bra a_3 \ket \over \si_o^3} + {s(s+1) \bra a_4 \ket \over \si_o^4} \cdots \bigg].
    \eeq
Differentiating and setting $s=0$ we get
    \beq
    {\zeta^{\prime}(0) \over \Om} &=& -{\si_o^2 \log{\si_o} \over 16 \pi^2} \bigg[
            {\bra a_0 \ket \over 2}  - {\bra a_1 \ket \over \si_o} + {\bra a_2 \ket \over \si_o^2}
            \bigg] + {\si_o^2 \over 16 \pi^2} \bigg[{3\bra a_0 \ket \over 4} - {\bra a_1 \ket
            \over \si_o} + \sum_{n=3}^\infty { (n-3)! \bra a_{n} \ket \over \si_o^{n}}
            \bigg] \cr
        &=& -{\si_o^2 \over 16 \pi^2} \bigg[ {\bra a_0 \ket \over 2} \log[\si_o
            e^{-3/2}] + {\bra a_1 \ket \over \si_o} (1-\log{\si_o}) + {\bra a_2 \ket
            \over \si_o^2} \log{\si_o}  - \sum_{n=3}^{\infty} {(n-3)! \bra a_n \ket \over \si_o^{n}}
            \bigg]
    \eeq
Inserting expressions for $\bra a_n \ket$ from
(\ref{e-list-of-avg-val-of-a_n}), we get a formula for
$\tr\log[-\grad^2 + \si] = -\zeta^{\prime}(0)$.
    \beq
        \tr\log[-\grad^2 + \si] &=& {\si_o^2 \Om \over 16\pi^2} \bigg[\half
            \log[{\si_o \over e^{3/2}}] + {\bra \vsi \ket \over \si_o}
            \log[{\si_o \over e}]
            + {\bra \vsi^2 \ket \over 2 \si_o} \log{\si_o} \cr && -
        \sum_{n=3}^\infty {\bra \vsi \grad^{2n-4} \vsi \ket \over n(n-1)(n-2) \si_o^n}
            \bigg] + {\cal O}(\vsi^3) \cr
        &=&  {\si_o^2 \Om \over 32\pi^2} \log[{\si_o \over
        e^{3/2}}] + \int d^4x \bigg[{\si_o \log[{\si_o \over e}] \over 16\pi^2}\vsi
            + {\log\si_o \over 32 \pi^2} \vsi^2
            \cr && -{1 \over 16\pi^2} \sum_{n=3}^\infty {\vsi (\grad^2)^{n-2}
            \vsi \over n(n-1)(n-2) (\si_o)^{n-2}} + {\cal O}(\vsi^3)
            \bigg].
    \eeq
The sum over powers of $\grad^2$ can be performed. Let $\Delta =
-{\grad^2 \over \si_o}$ and
    \beq
        \Pi(\Delta) = \sum_{n=1}^\infty {(-\Delta)^n
            \over n(n+1)(n+2)} = {\Delta(3 \Delta +2) -2(\Delta+1)^2
            \log{(1+\Delta)} \over 4 \Delta^2}.
    \label{e-inv-sigma-propagator}
    \eeq
$\Pi(\Delta)$ is analytic at $\Delta=0$,~~ $\Pi(\Delta) =
-{\Delta \over 6} + {\Delta^2 \over 24} - {\Delta^3 \over 60} +
\cdots$. For large \footnote{$\D = -{\grad^2 \over \si_o}$.
Note that $-\grad^2$ is a positive operator and $\si_o$ is not
a negative real number.} $\D$,
    \beq
    \Pi(\D) \to -\half \log \D + {3 \over 4} - {\log \D \over \D}
    + {\cal O}(\D^{-2}).
    \eeq
The final result is
    \beq
    \tr\log[-\grad^2 + \si] = {\si_o^2 \Om \over 32\pi^2}
        \log\bigg[{\si_o \over e^{3/2}}\bigg] + \int {d^4x \over 16\pi^2} \bigg[
        \si_o \log\bigg[{\si_o \over e}\bigg] \vsi + {\log{\si_o} \over 2} \vsi^2
            - \vsi \Pi(\Delta) \vsi + {\cal O}(\vsi^3) \bigg].
    \eeq
Here $\si(x) = \si_o + \vsi(x)$, $\Om$ is the volume of space-time
and $\Pi(\Delta)$ is defined above. This formula is valid for small
deviations of $\si$ from a constant background $\si_o$. The term
proportional to $\vsi$ vanishes if $\si_o$ is the average value of
$\si$. $\si$ need not be slowly varying.
We assumed that $\vsi$ approaches a constant as
$|x| \to \infty$ and that $\grad \vsi \to 0$ as $|x| \to \infty$.

{\fl \bf Remark:} If $\si$ is slowly varying, we ignore
terms with more than two derivatives to get
    \beq
    \tr \log[-\grad^2 + \si] = {\si_o^2 \Om \over 32 \pi^2} \log[{\si_o \over e^{3/2}}]
        + \int {d^4x \over 16\pi^2} \bigg[ \si_o \log[{\si_o \over e}]  \vsi
        + {\log \si_o \over 2} \vsi^2 - {\vsi \grad^2 \vsi \over 6 \si_o}  + {\cal O}(\vsi^3) \bigg]
    \eeq
where $\si = \si_o + \vsi$ and $\si_o$ is a constant.

\subsection{Scale anomaly $\zeta(0)$ for general backgrounds}
\label{a-scale-anomaly}

Though we only got an asymptotic series for $\zeta^{\prime}(0)$
around a constant background, we can get an exact closed-form
expression for its scale anomaly. Under a scale transformation $\si
\mapsto a^{2} \si$,
    \beq
        \zeta(s) \mapsto a^{-2s} \zeta(s)
        &\Rightarrow& \zeta^{\prime}(s) \mapsto - 2 \zeta(s)
            a^{-2s}\log a + a^{-2s} \zeta^{\prime}(s) \cr
        \zeta^{\prime}(0) &\mapsto& \zeta^{\prime}(0) - 2 \zeta(0) \log a
    \eeq
Now we use our result to find the scale anomaly $\zeta(0)$:
    \beq
         {\zeta(s) \over \Om} = {(\si_o)^{2-s} \over 16\pi^2} \bigg[{\bra a_0 \ket \over
            (s-1)(s-2)} + {\bra a_1 \ket \over (s-1) \si_o} + {\bra a_2 \ket \over \si_o^2} +
            {s \bra a_3 \ket \over \si_o^3} + {s(s+1) \bra a_4 \ket \over \si_o^4} \cdots
            \bigg].
    \eeq
Since all higher order terms are proportional to $s$, only
$a_0$, $a_1$ and $a_2$ contribute to $\zeta(0)$:
    \beq
        \zeta(0) = {\Om \si_o^2 \over 16 \pi^2} \bigg[
            {\bra a_0 \ket \over 2} - {\bra a_1 \ket \over \si_o} + {\bra a_2 \ket \over \si_o^2}
            \bigg] =
            {\Om \si_o^2 \over 16 \pi^2} \bigg[
            \half + {\bra \vsi \ket \over \si_o} + {\bra \vsi^2 \ket \over 2 \si_o^2}
            \bigg].
    \eeq

\section{Original potential $V(\phi)$ in zeta function regularization}
\label{a-back-to-V-legendre-transform}

The simplest way to describe the interactions of our model in
the large-$N$ limit is via the finite and scale-invariant
effective action $\Gamma_0(\B,\Si)$ of eqn.
(\ref{e-eff-ac-large-N-finite-and-scale-invariant}). Physical intuition and
approximation methods can be applied to $\Gamma$. The original potential $V(\phi^2/N)$
is {\em not} the effective potential, that honor goes to $\Gamma_0(B,\Si)$ for
constant backgrounds. The minima of $V(\phi^2/N)$ have no
physical significance. What is
more, $V(\phi^2/N)$ is {\em not} scale-free, it depends on a scale parameter $M$.
But, $M$ is not a parameter of the theory, it is
canceled by scale `anomalies' from quantum fluctuations.
Moreover, $W(\si)$ is divergent, so $V(\eta)$ is not strictly
defined independent of a regularization scheme. Despite all these warnings,
many physicists wish to know what $V(\eta)$ is, so we find
the $V(\eta)$ that corresponds to $W(\si)$ obtained in zeta
function regularization. In the large-$N$ limit, we find that
for a constant background field $\si$, $V(\eta)$ grows as
$\eta^2 / \log \eta$ for large $\eta = \phi^2/N$. We haven't
yet determined its behavior for small $\eta$.

Recall that $e^{-N V(\eta(x))}$ is the inverse Laplace transform of $e^{-N
W(\si(x))}$ for each $x$:
    \beq
    \int_{\cal C} \fr{d\si}{2\pi i} e^{-(N/2 \hbar) (W(\si) + \si
    \eta)} = e^{-(N/2\hbar)V(\eta)}.
    \eeq
We found $W(\si)$ only at $N =\infty$, so it makes sense to invert
the Laplace transform for large-$N$. We do this here for
constant $\si$, for which we have found\footnote{We
should have a mass scale $M$ to set the scale for the logarithms. We set
$M=1$ in this section.} $W(\si)$ exactly in (\ref{e-W_0-for-constant-bkgrnd}):
    \beq
    W(\si) + \eta \si = \eta \si - m^2 \si - {\hbar \si^2 \over 32
    \pi^2} \log(\tl \la \si)
    {\rm ~~~where~~~~}  \tilde \la = e^{-[32 \pi^2 \la \hbar^{-1} +
    3/2]}.
    \eeq
The branch cut of $W(\si)$ implies ${\cal C}$ runs from $-i \infty$ to $i \infty$
avoiding the negative real $\si$ axis.
Putting $\si = u + iv$ for $u, v \in {\bf R}$, the real and
imaginary parts of $W(\si) + \eta \si = \vphi + i \psi$ are ($\hbar =1$)
    \beq
    \vphi &=& (\eta -m^2)u - {(u^2-v^2) \log{\tl \la \sqrt{u^2 + v^2}} \over 32
    \pi^2}
        + {uv \arctan(v/u) \over 16 \pi^2} \cr
    \psi &=& (\eta -m^2)v - {(u^2-v^2) \arctan(v/u) \over 32
    \pi^2} - {uv \log(\tl \la \sqrt{u^2 + v^2}) \over 16
    \pi^2}.
    \eeq
We wee that $\Re (W(\si) + \eta \si) \to \infty$ as $v \to \pm
\infty$ for any $u \geq 0$. So the integrand vanishes along the
lines $u \pm i \infty$ for any $u$. Thus, the end points of
${\cal C}$ can be moved from $\pm i \infty$ to $\pm i \infty +
u_{\pm}$ for any real $u_{\pm}$ without altering the integral.
Simply put, it does not matter along which longitude ${\cal C}$
leaves the south pole or along which longitude it approaches
the north pole.

The general strategy for estimating such integrals is as
follows \cite{bender-orszag}. $W(\si) + \eta \si$ is in general
complex on ${\cal C}$. Its imaginary part $\psi$
will lead to a highly oscillatory integral as $N \to \infty$
and make it difficult to estimate. The trick is to use
analyticity of $W(\si) + \eta \si$ to deform the contour to a
(union of) contour(s) along which $\Im(W(\si) + \eta \si)$ is
constant and other contours where the integrand vanishes. Such
contours are called constant phase contours and coincide with
the steepest contours, those along which the absolute value of
the integrand changes fastest. If ${\cal C}$ is such a contour
(assumed to be a single one for simplicity), then
    \beq
    \int_{\cal C} \fr{d\si}{2\pi i} e^{-(N/2 \hbar) (W(\si) + \si
    \eta)} = e^{-{N i \over 2 \hbar} \Im(W(\si) + \si \eta)}
    \int_{\cal C} \fr{d\si}{2\pi i} e^{-(N/2 \hbar) \Re(W(\si) + \si
    \eta)}.
    \eeq
Now we have eliminated the oscillating phase and for
large-$N$, the asymptotic behavior is
determined by the local minima of $\vphi = \Re(W(\si) + \si
\eta))$ along ${\cal C}$. Since $\vphi$ diverges at the end
points of ${\cal C}$, local minima must occur at interior
points of ${\cal C}$. Moreover, there must be an odd number
$1,3,5,\cdots$ of such local minima along ${\cal C}$. At any
one, the directional derivatives of both $\vphi$ and $\psi$
vanish in the direction tangent to the curve. Since $\vphi + i
\psi$ is analytic, it follows that these local minima of
$\vphi$ are saddle points, i.e. $\pdr_\si(W(\si) + \si
\eta)=0$, where two or more steepest curves intersect. Not all
saddle points of $W(\si) + \si \eta$ need lie on ${\cal C}$ and
those that don't will not contribute to the asymptotic behavior
of the integral.

Suppose $\si = \si_s(\eta)$ is the only saddle point along the
constant phase contour ${\cal C}$. The integrand attains a
local maximum at $\si_s$ along ${\cal C}$ and decays
exponentially in either direction away from $\si_s$. The
contour can be approximated by a straight line tangent to $\cal
C $ at $\si_s$ and of length $\eps$ on either side. $\vphi(\si)
= \Re(W + \eta \si)$ is approximated by its quadratic Taylor
polynomial around $\si_s$, whose linear term vanishes. Now we
let $\eps \to \infty$. $\vphi(\si_s)$ gives the leading
contribution while the quadratic term in its Taylor series
gives a gaussian integral proportional to $\ov{\sqrt{N}}$. So
    \beq
    e^{-{N \over 2} V(\eta)} &=& e^{-{iN \over 2} \psi(\si_s)}
        e^{-{N \over 2} \vphi(\si_s)}
        {1 \over 2\pi i} {\cal O}(\ov{\sqrt{N}}), \cr
    \implies ~~~ V(\eta) &=& \vphi(\si_s) + i \psi(\si_s) + {\cal O} \bigg({\log N \over N}
    \bigg).
    \eeq
If there is more than one saddle point on $\cal C$, we add up
their contributions in this formula for $V(\eta)$. Moreover, if
the saddle point $\si_s$ is on the real axis, then $\psi(\si_s)
= 0$ does not contribute.

So our job is to find a convenient constant phase contour and
identify the saddle points on it. In practice, we look for
saddle points and then a suitable contour.  The saddle point
condition for $W(\si) + \si \eta$ for given $\eta, m^2$ and
$\la$ is
    \beq
    {16 \pi^2 \over \hbar} (\eta -m^2) = \si \log(\tl \la \si
    \sqrt{e}).
    \label{e-saddle-pt-eqn}
    \eeq
Taking imaginary and real parts, it is a pair of transcendental
equations (for $\tl \la =1$ and $\hbar =1$)
    \beq
    {v \over 2} + {v \over 2} \log{(u^2 + v^2)} &=&
        - u \arctan(v/u) {\rm ~~~~and} \cr
    16 \pi^2 (\eta - m^2) &=& {u \over 2} + {u \over 2} \log{(u^2 +
        v^2)} - v \arctan(v/u).
    \eeq
The fundamental domain for the arctangent is taken as $-\pi <
\arctan < \pi$. We must solve for $\si = u + iv$ assuming it is
off the negative real axis. Any $u>0$ for $v=0$ satisfies the
first condition (i.e. saddle points can lie on the positive
real $\si$ axis), but there are other possibilities. The
imaginary part of the saddle point condition is also satisfied
along a curve (found numerically) in the $u-v$ plane that
encircles the origin and is symmetric under reflections about
either axis\footnote{This follows from the even and oddness of
the condition in $u$ and $v$ respectively.} and lies within the
rectangle\footnote{The limiting values are obtained by solving
the first saddle point condition for small $v$ and $u$
respectively.} $|u| \leq e^{-3/2}, |v| \leq \ov{\sqrt{e}}$.
However, the second saddle point condition is satisfied on this
curve only for a limited range of values of $\eta -m^2$, namely
$m_c^2 \geq \eta - m^2 \geq -\ov{32 \pi \sqrt{e}}$ for
$-e^{-3/2} \leq u \leq 0$ and $-\ov{32 \pi \sqrt{e}} \leq \eta
- m^2 \leq -m_c^2$ for $0 \leq u \leq e^{-3/2}$ where $m_c^2 =
{\hbar e^{-3/2} \over 16 \pi^2 \tl \la}$. For $\eta - m^2$ in
this range, saddle points could occur on the above mentioned
curve as well as the positive real $\si$ axis, making their
analysis more involved. For now, we set aside the behavior of
$V(\eta)$ for small\footnote{Recall that we are working in
units where the scale parameter $M=1$.} $\eta$, i.e. $\eta -
m^2 \leq m_c^2$. For $\eta - m^2 \geq m_c^2$ the only possible
saddle points are located on the positive real $\si$ axis
$(v=0)$. In this case, the positions of the saddle points are
given by the solutions to equation (\ref{e-saddle-pt-eqn})
where $\si = u$ is real. Since we assumed $\eta - m^2 > m_c^2$,
the LHS is in particular positive, and there is a unique
solution $\si_s$ which can be found recursively
    \beq
    \si_s = {\tl \eta \over \log(\tl \la \sqrt{e} \si_s)} = {\tl \eta \over
        \log\bigg( {\tl \la \sqrt{e} \tl \eta \over \log (\tl \la \sqrt{e} \si_s)} \bigg)}
        = {\tl \eta \over \log(\tl \la \sqrt{e} \tl \eta) - \log \log (\tl \la \sqrt{e} \si_s)}
        = \cdots
    \eeq
where $\tl \eta = 16 \pi^2 (\eta - m^2)/\hbar$. Thus, for
sufficiently large $\eta - m^2$, there is only a single saddle
point $\si_s$. Moreover, we have checked numerically that there
is a constant phase (actually zero phase, since
$\psi(\si_s)=0$) contour that starts at the south pole, passes
through $\si_s$ and goes to the north pole. For large $\eta -
m^2$, the leading approximation for the position of the saddle
point is (with $\bar \la = 16 \pi^2 \sqrt{e} \tilde \la /
\hbar$)
    \beq
    \si_s \to {16 \pi^2 (\eta - m^2) \over \hbar \log[\bar \la (\eta - m^2)]}
    {\rm ~~ ~~~as~~~~} \eta -m^2 \to \infty.
    \eeq
The original potential is given by $V(\eta) =  W(\si_s(\eta)) +
\eta \si_s(\eta) + {\cal O}(\log N/N)$. Using the saddle point
equation (\ref{e-saddle-pt-eqn}), we simplify this to
    \beq
        V(\eta) = W(\si_s) + \eta \si_s = \fr{\hbar \si_s^2}{64 \pi^2} + \half \si_s
        (\eta - m^2).
    \eeq
For large $\eta -m^2$ we get\footnote{Recall that $\bar \la =
16 \pi^2 \sqrt{e} \tilde \la / \hbar$ and $\tl \la = e^{-(32
\pi^2 \la/\hbar + 3/2 )}$}
    \beq
        V(\eta) \to {8 \pi^2 \over \hbar} {(\eta -m^2)^2 \over \log[\bar\la (\eta - m^2)]} \bigg[1
        + \ov{2 \log[\bar\la (\eta -m^2)]} \bigg] {\rm ~~ as~~} \eta - m^2 \to
        \infty
    \eeq
Recalling that $\eta = {\phi^2 \over N}$, we see that up to a
multiplicative factor, $V(\phi^2/N) \sim {(\phi^4/N^2) \over
\log(\phi^2/N)}$ for large $\phi^2/N$ and fixed $m$. Thus, the
original potential grows logarithmically slower that a quartic
potential in the large-$N$ limit. It would be interesting to
find the behavior of $V(\eta)$ for small $\eta$. This requires
a careful study of saddle points and constant phase contours
for small $\eta$.



\end{document}